%% file: main.tex
\documentclass[journal]{IEEEtran}  
\usepackage{amsmath,amsfonts}
\usepackage{algorithmic}
\usepackage{algorithm}
\usepackage{array}
\usepackage[caption=false,font=normalsize,labelfont=sf,textfont=sf]{subfig}
\usepackage{textcomp}
\usepackage{stfloats}
\usepackage[hyphens]{url}
\usepackage{verbatim}
\usepackage{graphicx}
\usepackage{cite}
\usepackage{multirow}
\usepackage{multicol}
\usepackage{xcolor}
\usepackage{hyperref}

\newcommand{\norm}[1]{\left\lVert#1\right\rVert}

\newcommand{\highlight}[1]{{#1}}

\newcolumntype{P}[1]{>{\centering\arraybackslash}p{#1}}
\mathchardef\mhyphen="2D

\begin{document}

\title{Joint Quality Assessment and Example-Guided Image Processing by Disentangling Picture Appearance from Content}

\author{Abhinau K. Venkataramanan, Cosmin Stejerean, Ioannis Katsavounidis,\\ Hassene Tmar, and Alan C. Bovik~\IEEEmembership{Life Fellow,~IEEE}
\thanks{This research was sponsored by a grant from Meta Video Infrastructure, and by grant number 2019844 for the National Science Foundation AI Institute for Foundations of Machine Learning (IFML).}}

\markboth{Journal of \LaTeX\ Class Files,~Vol.~14, No.~8, August~2021}%
{Venkataramanan \MakeLowercase{\textit{et al.}}: One Transform to Compute Them All}


\maketitle

\begin{abstract}
The deep learning revolution has strongly impacted low-level image processing tasks such as style/domain transfer, enhancement/restoration, and visual quality assessments. Despite often being treated separately, the aforementioned tasks share a common theme of understanding, editing, or enhancing the appearance of input images without modifying the underlying content. We leverage this observation to develop a novel disentangled representation learning method that decomposes inputs into content and appearance features. The model is trained in a self-supervised manner and we use the learned features to develop a new quality prediction model named DisQUE. We demonstrate through extensive evaluations that DisQUE achieves state-of-the-art accuracy across quality prediction tasks and distortion types. Moreover, we demonstrate that the same features may also be used for image processing tasks such as HDR tone mapping, where the desired output characteristics may be tuned using example input-output pairs.
\end{abstract}


\input{sections/introduction}
\input{sections/background}
\input{sections/disque}
\input{sections/datasets}
\input{sections/evaluation}
\input{sections/discussion}
\input{sections/conclusion}

\bibliographystyle{IEEEtran}
\bibliography{refs}

\newpage

 





\end{document}

%% file: sections/introduction.tex
\section{Introduction}
\label{sec:introduction}

Recent years have witnessed an explosion in the amount of image and video content being shared over the internet. These images and videos are captured, often by uncertain hands, using cameras of various capabilities that may introduce distortions such as blur, noise, under/overexposure, etc. Following capture, they are commonly subjected to distortions such as compression, scaling, and brightness or contrast distortions during the transmission and display processes. Moreover, images and videos may also be edited by artists to modify their appearance, by making images brighter or darker, changing colors and color saturation, boosting contrast, etc.

At the same time, more sophisticated imaging and display modalities such as high dynamic range (HDR), high frame rate (HFR), and immersive media are rapidly growing. In particular, HDR enables the capture and representation of a wider range of brightnesses and colors, thereby enabling a more realistic reproduction of natural scenes as compared to legacy Standard Dynamic Range (SDR) imaging systems. However, a substantial portion of existing displays are not capable of displaying brightnesses above 1000 nits, which is essential for HDR \cite{ref:hdr_tv_survey}. So, HDR images and videos must be down-converted to SDR using a process called tone-mapping, so that they may be displayed on legacy displays. Although several algorithms have been proposed to automatically perform tone mapping \cite{ref:itu21} \cite{ref:hable}, color grading by human experts remains the gold standard.

Therefore, in both of the aforementioned scenarios, corresponding to the handling of SDR and HDR images/videos, two tasks must be effectively conducted to reliably transmit high-quality videos to consumers. First, objective models are needed that predict subjective opinions regarding the visual quality of images and videos. Such models may be used to control the quality of ingested content on social media websites and identify poor-quality content, which may affect downstream recommendation decisions. Quality models have also been used extensively to optimize processing parameters such as compression and resolution while trading off storage and transmission costs against perceptual quality \cite{ref:dyn_optimizer}.

Secondly, to control the quality of streamed content, image processing methods are also needed that can enhance specified aspects of images, such as brightness, contrast, color, etc. Typically, such fine-grained editing requires the use of specialized algorithms that provide ``tunable knobs'' corresponding to various image features. Indeed, this approach has led to the development of many tone-mapping algorithms that contain parameters to control aspects of the appearance of tone-mapped images. For example, the photographic tone reproduction method \cite{ref:reinhard02}, which we refer to as ``Reinhard02'' here, uses a ``desaturation'' parameter to correct oversaturated colors. However, a human expert who is performing tone-mapping or evaluating its quality does not target a ``desaturation level.'' Rather, a colorist tunes image properties manually to achieve a desired ``look,'' which may depend on the colorist's experience and preferences, and their perception of consumers' demands. We posit that such specifications are best described using examples, rather than analytical metrics. This motivates the task of ``example-guided'' image processing, in general, or tone-mapping in particular. More examples of tone-mapping methods and their tunable parameters are provided in Section \ref{sec:datasets_hdr}.

\begin{figure}[t]
    \centering
    \includegraphics[width=\linewidth]{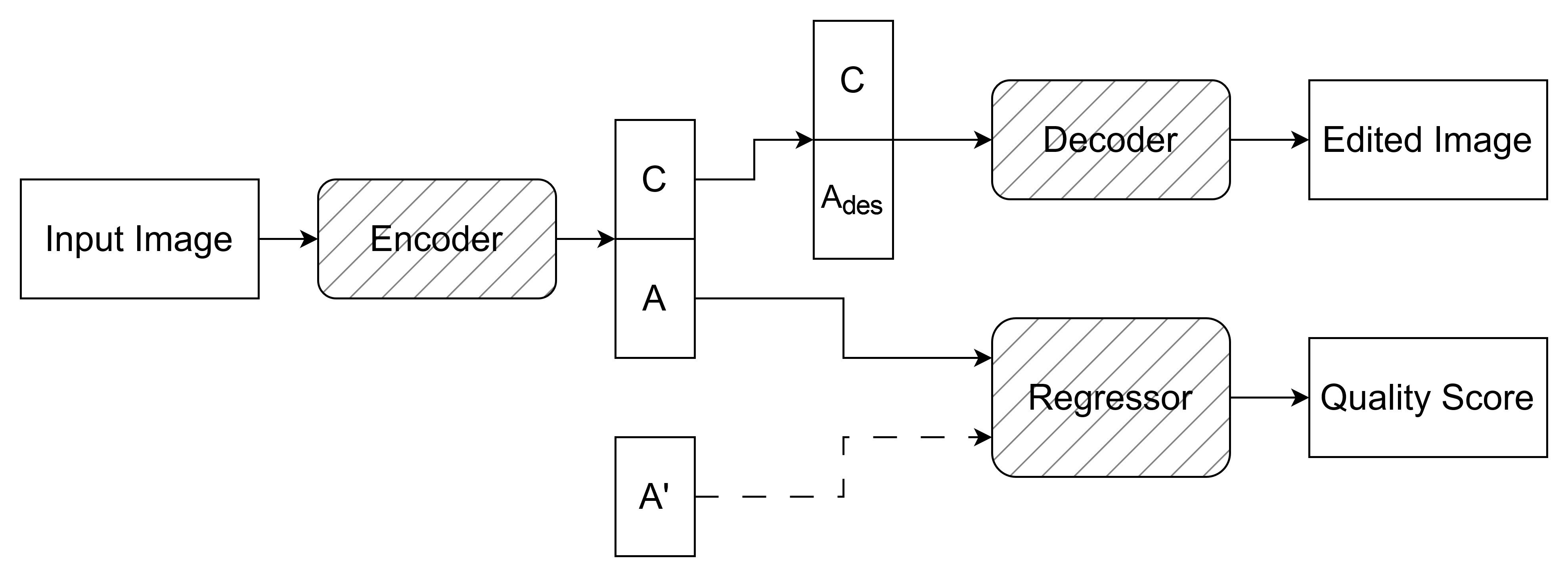}
    \caption{Performing both quality prediction and image processing using the same disentangled representation learning model.}
    \label{fig:disrep_proposal}
\end{figure}

Here, we propose a Disentangled Representation Learning (DRL) framework to create a deep neural network model that can be used to tackle both image quality prediction and image processing tasks simultaneously. The general framework for using a common deep model for both quality assessment and image processing tasks is illustrated in Fig. \ref{fig:disrep_proposal}.

First, an input image is decomposed into two feature sets, each describing the ``image content'' and ``image appearance.'' During quality modeling, the appearance feature is compared against a reference appearance feature, to predict subjective quality. For example, when measuring the visual quality of a tone-mapped HDR video frame, its appearance feature is compared against the appearance feature of the source HDR video frame. We call this model the \textbf{Disentangled Quality Evaluator (DisQUE)}.

The input image may then be edited or enhanced by modifying the appearance feature of the input image to a ``desired appearance feature,'' followed by reconstructing it using a decoder. We obtain the desired appearance feature using a pair of input images that are used as an example of (in this case) the desired tone-mapping behavior. We term this \textbf{``example-guided tone mapping'' (EGTM).}

The remainder of this paper is organized as follows. In Section \ref{sec:background}, we discuss relevant prior work in the fields of visual quality assessment (VQA) and DRL and explain the novelty of our proposed model. Section \ref{sec:disque} provides a detailed description of our proposed DisQUE model, including the learning objective, deep neural network architecture, and feature extraction protocol for quality prediction. In Section \ref{sec:datasets}, we describe the training and evaluation datasets corresponding to the two domains in which we evaluate DisQUE. Specifically, we describe the datasets used by DisQUE to predict the quality of tone-mapped and compressed HDR videos, and to predict the quality of SDR images. We present training details in Section \ref{sec:evaluation_training} and the results of quality modeling experiments in Section \ref{sec:evaluation_quality}. Furthermore, we demonstrate the ability of the DRL model to perform example-guided HDR tone-mapping in Section \ref{sec:evaluation_tonemapping}. Finally, we present a summary of our findings and identify avenues for future work in Section \ref{sec:conclusion}.

%% file: sections/background.tex
\section{Background and Novelty}
\label{sec:background}

\input{sections/background_subsections/quality_assessment}
\input{sections/background_subsections/disrep_learning}
\input{sections/background_subsections/novelty}

%% file: sections/background_subsections/quality_assessment.tex
\subsection{Visual Quality Assessment}
\label{sec:background_quality_assessment}
Objective models of visual quality may be broadly classified into ``classical'' (or hand-crafted) or ``deep'' (data-driven deep networks) methods. Full-reference (FR) quality models compare ``distorted'' test pictures/videos against their ``pristine'' reference counterparts to predict their visual quality. Models like SSIM \cite{ref:ssim}, VIF \cite{ref:vif}, and ST-RRED \cite{ref:strred} are examples of general-purpose classical FR quality models. By contrast, models such as DLM \cite{ref:dlm}, VMAF \cite{ref:vmaf}, and FUNQUE \cite{ref:funque}\cite{ref:funque_plus} are task-specific models designed to predict the quality of scaled and compressed videos.

FR models targeting similar applications have also been developed for HDR pictures and videos. Examples of such quality models include HDRMAX-VMAF \cite{ref:hdrmax_vmaf} and HDR-FUNQUE+ \cite{ref:hdr_funque_plus}. In addition, quality models such as TMQI \cite{ref:tmqi}, FSITM \cite{ref:fsitm}, and Cut-FUNQUE \cite{ref:cut_funque} compare HDR and SDR pictures and videos to assess the quality of HDR tone mapping.

Deep FR quality modeling may be performed using deep networks pre-trained on large datasets such as ImageNet. For example, LPIPS \cite{ref:lpips}, DISTS \cite{ref:dists}, and DeepWSD \cite{ref:deep_wsd} all utilize ImageNet-pretrained models.

When pristine reference content is not available, No-reference (NR) quality models are employed. Examples of classical NR models include BRISQUE \cite{ref:brisque}, DIIVINE \cite{ref:diivine}, TLVQM \cite{ref:tlvqm}, HIGRADE \cite{ref:higrade}, and ChipQA \cite{ref:chipqa}. Deep NR models may be trained either in a supervised or a self-supervised manner. Examples of supervised deep NR models include CNN-based models such as PaQ-2-PiQ \cite{ref:paq2piq}, Patch-VQ \cite{ref:patchvq}, and QFM-IQM \cite{ref:qfm_iqm}, and transformer-based models such as MUSIQ \cite{ref:musiq}, RKIQT \cite{ref:rkiqt}, LoDA \cite{ref:loda}, and SaTQa. 

Recently, a number of high-performing self-supervised NR models have been introduced, including CONTRIQUE \cite{ref:contrique}, Re-IQA \cite{ref:reiqa}, and ConViQT \cite{ref:conviqt}, and  may also be used for NR quality prediction. All three self-supervised methods utilize ResNet-50 backbones and contrastive learning techniques such as SimCLR \cite{ref:simclr} and MoCo \cite{ref:moco} \cite{ref:mocov2} to learn quality-aware representations. The predicted features from test images/videos may be used for NR quality modeling, while the differences in predicted features between the reference and test images may be used for FR quality modeling.

%% file: sections/background_subsections/disrep_learning.tex
\subsection{Disentangled Representation Learning}
\label{sec:background_disrep_learning}
Disentangled representation learning (DRL) refers to representation learning techniques that impose a notion of independence between subsets of the learned features. A survey of DRL methods and various taxonomic classifications are provided in \cite{ref:drl_survey}. Consider a network learning a vector of features. The disentanglement condition may be applied to each dimension of the feature vector, which reflects the assumption that each dimension encodes one generative factor of the data distribution being modeled. Examples of dimension-wise disentangling include variational autoencoder (VAE) methods such as FactorVAE \cite{ref:factor_vae} and \(\beta\)-TCVAE \cite{ref:beta_tcvae}, generative adversarial network (GAN) methods such as InfoGan \cite{ref:info_gan}, PS-SC GAN \cite{ref:ps_sc_gan} and OroJaR GAN \cite{ref:orojar}, and Barlow Twins \cite{ref:barlow}, which is a self-supervised representation learning method.

Rather than disentangling feature vectors dimension-wise, their subsets may be disentangled to separate specific aspects of the data distribution. For example, DR-GAN \cite{ref:dr_gan} separates face and pose information to conduct pose-invariant face retrieval, while MAP-IVR \cite{ref:map_ivr} disentangles content and motion information from videos to conduct image-to-video retrieval. Feature subsets may be disentangled using cosine distances \cite{ref:map_ivr}, minimizing correlations \cite{ref:barlow}, or by minimizing mutual information using techniques such as CLUB \cite{ref:club_mi} or adversarial losses \cite{ref:adv_disrep}.

Two key applications of DRL to image processing that are relevant here are style transfer and domain adaptation. The goal of DRL in such tasks is to decompose an image into a feature set that is common across domains, typically encoding ``content,'' and one that is domain-specific, typically encoding ``style.'' Examples of such methods include DRIT++ \cite{ref:drit} and \cite{ref:disrep_domain}.

Efforts have also been made to combine image restoration and image quality assessment tasks, both with and without disentangled representations. QAIRN \cite{ref:qairn} uses a residual attention mechanism to gate encoder and decoder signals in image restoration networks. Due to the gating effect, the attention maps in QAIRN have been shown to learn local quality-aware feature maps. QD-Net \cite{ref:qdnet} uses disentangled representations to perform no-reference (NR) quality prediction and enhancement of tone-mapped HDR images. QD-Net is trained in a supervised manner using ground-truth subjective ratings, while the enhancement network is trained using predefined enhancement targets. Alhough the ``amount of enhancement'' may be varied, the exact nature of the enhancement, such as improving color, brightness, or contrast, cannot be controlled. Therefore, if the desired enhancement targets change, for example, due to changes in consumer preferences, the enhancement network must be retrained. 

Similar to QD-Net, DRIQA \cite{ref:driqa} targets joint supervised quality assessment and restoration of SDR images. In this work, the ``content encoder'' extracts representations that do not contain distortion information. Therefore, the output of the content encoder is used directly for restoration, while the appearance encoder captures only distortion-related information. The outputs of the distortion encoder are used to augment a Siamese network \cite{ref:siamese} designed for quality assessment. Once again, the restoration targets are fixed and the decoder must be retrained if the restoration behavior is to be modified.

%% file: sections/background_subsections/novelty.tex
\subsection{Novelty of the Proposed DRL Method}
\label{sec:background_novelty}
Our proposed DRL method differs from prior work in the following key aspects. First, prior work on disentangled domain transfer either uses different encoders/decoders for each domain \cite{ref:disrep_domain} and/or categorical inputs to specify the target domain \cite{ref:drit} for multi-domain adaptation. By contrast, we use a \textbf{fixed pair of networks to disentangle features \highlight{without the need for a domain label,}} and the predicted appearance features are used directly to specify the target domain for image processing. \highlight{This property of our proposed method enables its use for quality assessment. The ``domain'' in style transfer controls the visual appearance of images, the characterization of which is the goal of visual quality assessment. However, since existing work requires choosing a different encoder for each domain or providing a domain label to the encoder, those encoders cannot themselves be used to obtain a representation of the domain, i.e., quality.}

Secondly, both QD-Net and DRIQA solve restoration tasks using pre-defined restoration targets. We instead adopt an \textbf{``example-guided image processing'' (EGIP)} framework that learns general appearance-related representations at training time. During inference, the desired processing behavior is expressed using a pair of example images that include a source image and its processed version. The DRL model infers the desired transform from the example and applies it to the input source image to be processed.

Thirdly, we propose a novel method for adapting images across domains called \textbf{``appearance mixing,''} as an alternative to ``appearance replacement'' methods used \highlight{in the style transfer literature} \cite{ref:drit} \cite{ref:disrep_domain} \cite{ref:qdnet}. We observed that appearance replacement led to inaccurate adaptation across domains due to the presence of ``confounding appearance features'' (CAFs) in the source image. We demonstrate the effects of CAFs in Section \ref{sec:evaluation_tonemapping} and show that they may be mitigated by using appearance mixing.

Finally, both QD-Net and DRIQA-NR were trained in a supervised manner using ground-truth subjective quality scores. By contrast, DisQUE is trained in a \textbf{self-supervised} manner without the need for subjectively annotated data.

%% file: sections/disque.tex
\section{DisQUE}
\label{sec:disque}
Here, we describe our proposed disentangled representation learning algorithm, which we use to develop DisQUE. The goal of the learning algorithm is to decompose an input image into its ``content'' and ``appearance'' components. In prior models such as ReIQA \cite{ref:reiqa}, ``content'' features are extracted to identify semantic content, such as objects in the image. To achieve high object detection accuracy, these models are designed to be robust to small changes in structure and orientation. By contrast, small changes in structure are visible to the human eye and perceptually important, as evidenced by advancements in restoration tasks such as image deblurring.

Here, we interpret the content to be the ``high-resolution'' intrinsic structure of the image, which acts as a scaffolding that is modulated by ``appearance.'' Appearance, on the other hand, are properties that vary slowly across an image and include aspects such as color, contrast, brightness, and sharpness. Modeling content and appearance under these assumptions is similar to the decomposition of scenes into reflectance and illumination components \cite{ref:intrinsics}.

\input{sections/disque_subsections/learning_objective}
\input{sections/disque_subsections/architecture}
\input{sections/disque_subsections/quality_assessment}

%% file: sections/disque_subsections/learning_objective.tex
\subsection{Learning Objective}
\label{sec:disque_learning}
\begin{figure*}[t]
    \centering
    \includegraphics[width=0.75\linewidth]{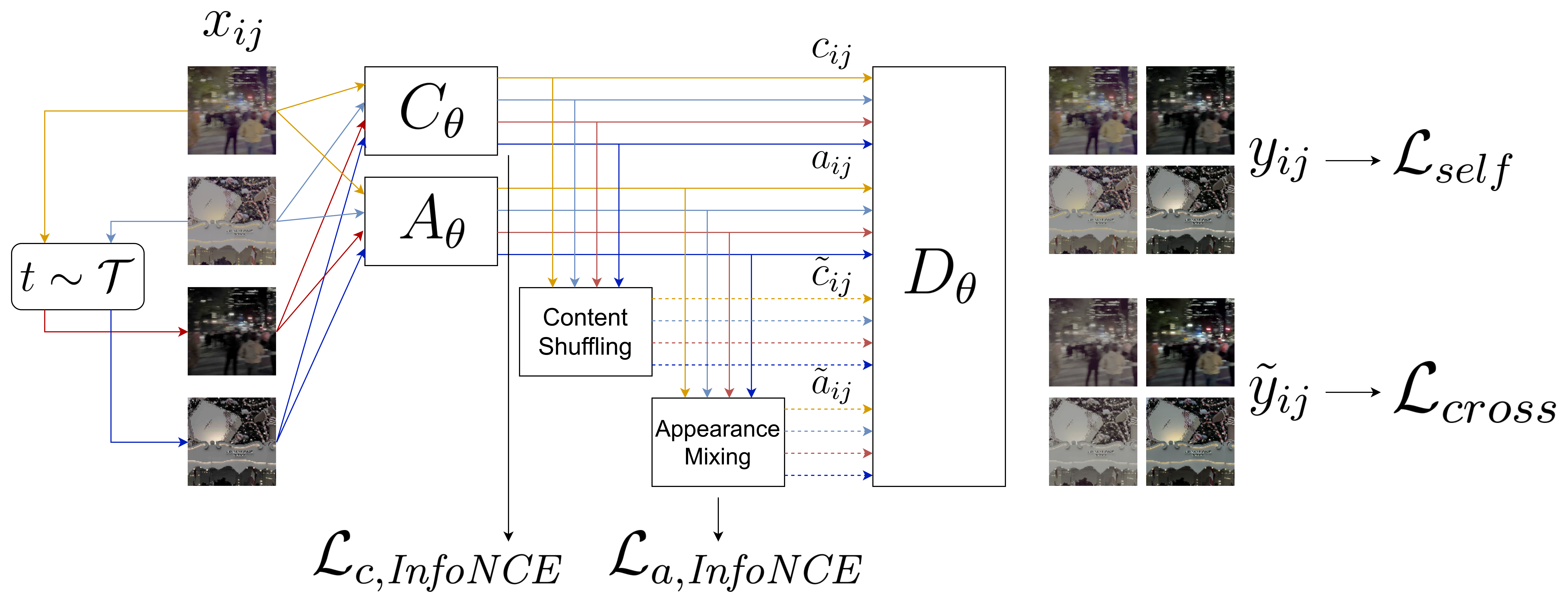}
    \caption{Visualizing the disentangled representation learning objective.}
    \label{fig:disque_training}
\end{figure*}
Consider a dataset of images \(\mathcal{X} = \{x_i\}\) and a bank of image transforms \(\mathcal{T} = \{T_j\}\) that alter one or more aspects of the appearance of input images. Examples of such transforms include blurring, brightening, compression, color changes, or tone-mapping operators for HDR. A list of transforms commonly used in HDR and SDR experiments is presented in Section \ref{sec:datasets}. We sample two image patches \(x_1, x_2 \sim \mathcal{X}\) and a transform \(t \sim \mathcal{T}\). Applying the transform to both images, we obtain two views of each image:
\begin{equation}
    x_{11} = x_1, \quad x_{12} = t\left(x_1\right), \quad x_{21} = x_2, \quad x_{22} = t\left(x_2\right).
    \label{eq:x_aug}
\end{equation}

\highlight{In the style transfer literature, the transform \(t\) may denote a particular target domain, such as converting natural images into cartoons or paintings in a particular style. Here, \(t\) is sampled from a rich class of transformations that affect the low-level characteristics of images, such as sharpness, blur, contrast, color saturation, etc., applied to varying degrees. In our formulation, \(x_{11}, x_{21}\) always denote two pristine SDR/HDR images. The transform \(t\) may be chosen such that \(x_{12}, x_{22}\) denote the output of applying Gaussian blur with a standard deviation of 0.5, or increasing the contrast by 20\%, or adding white Gaussian noise with strength 0.2. The goal of the learning objective is to disentangle the type and amount of a distortion from the underlying content of the pristine images.}

Let \(C_\theta\) and \(A_\theta\) denote two parameterized encoders that map input images to their content and appearance features respectively:
\begin{equation}
    c_{ij} = C_\theta(x_{ij}), \quad a_{ij} = A_\theta(x_{ij}).
\end{equation}
Finally, let \(D_\theta\) denote a parameterized decoder that maps content and appearance features to images. When content and appearance features extracted from an image are reconstructed, we expect to recover the input image. We term this the \textbf{self-reconstruction} objective. The reconstruction loss is a weighted sum of pixel-domain and frequency-domain losses, as used in MAXIM \cite{ref:maxim}. That is, let
\begin{equation}
    y_{ij} = D_\theta\left(c_{ij}, a_{ij}\right),
\end{equation}
\begin{equation}
    \mathcal{L}_{self} = \sum\limits_{ij} \mathcal{L}_{char}(x_{ij}, y_{ij}) + \lambda_f \mathcal{L}_{freq}(x_{ij}, y_{ij}),
\end{equation}
where \(\mathcal{L}_{char}\) denotes the Charbonnier loss
\begin{equation}
    \mathcal{L}_{char}(x, y) = \sqrt{\norm{x - y}^2 + \epsilon^2}
\end{equation}
and \(\mathcal{L}_{freq}\) denotes the frequency loss, which uses the discrete Fourier transform, denoted by \(\mathcal{F}\)
\begin{equation}
    \mathcal{L}_{freq}(x, y) = \norm{\mathcal{F}\left\{x\right\} - \mathcal{F}\left\{y\right\}}_1.
\end{equation}

The main contribution of our proposed disentangled representation learning algorithm, which enables the separation of content and appearance features, is the \textbf{\highlight{appearance-mixing} cross-reconstruction objective}. \highlight{It is important to note that the cross-reconstruction objective has been extensively used in style transfer applications \cite{ref:drit} \cite{ref:disrep_domain}. Generally, a cross-reconstruction objective refers to combining domain-agnostic features of an image with domain-specific features from another image to transfer the style/domain of the first image. The novelty of DisQUE method is in the method used to infer the domain-specific feature, which we term ``appearance mixing.'' A more detailed formulation of cross-reconstruction and the proposed appearance-mixing method follows.}

\highlight{Using the previously defined notation,} the goal of cross-reconstruction is to predict an image \(x_{ij}\) using features from images other than \(x_{ij}\). For example, suppose we wish to predict \(x_{12}\) \highlight{from the pristine image \(x_{11}\) and the distorted image \(x_{22}\)}. Since \(c_{11}\) encodes the content in image 1 and \(a_{22}\) encodes appearance after applying transformation \(t\), one may predict \(x_{12}\) as \(D_\theta\left(c_{11}, a_{22}\right)\). Such a cross-reconstruction method has been used in prior work, such as DRIT \cite{ref:drit}, to disentangle content and appearance. \highlight{We refer to this as ``appearance replacement,'' since the appearance feature \(a_{11}\) of the image \(x_{11}\) is replaced by \(a_{22}\) of \(x_{22}\) to predict \(x_{12}\).}

However, the appearance feature \(a_{22}\) includes information not only about the effect of \(t\), but also of the source image \(x_2\). Such ``confounding'' appearance features (CAFs) may be transferred if cross-reconstruction is performed in this manner. For example, if \(x_1\) is a picture of a green field and \(x_2\) is that of a yellow flower, \(D_\theta\left(c_{11}, a_{22}\right)\) may yield a field with a yellow hue.

To remove the effect of CAFs, we adopt a novel \textbf{``appearance mixing''} method. We first note that \highlight{since \(x_{i2} = t(x_{i1})\),} the difference between \(x_{i1}\) and \(x_{i2}\) is only the effect of the transform \(t\). Therefore, \(\Delta a_i = a_{i2} - a_{i1}\) captures the effect of \(t\) while eliminating CAFs from image \(x_i\). This difference is then ``mixed into'' other appearance features to add or remove the effect of \(t\) and yield cross-reconstructed images. Hence, the crossed appearance features after undergoing appearance mixing are
\begin{align}
    \Tilde{a}_{11} = a_{12} - \Delta a_2, \nonumber \\
    \Tilde{a}_{12} = a_{11} + \Delta a_2, \nonumber \\
    \Tilde{a}_{21} = a_{22} - \Delta a_1, \nonumber \\
    \Tilde{a}_{22} = a_{21} + \Delta a_1.
\end{align}

\highlight{The first two transformations may, respectively, be interpreted as follows:}
\begin{itemize}
    \item \highlight{The appearance of the first pristine image is predicted by removing the effect of the transform \(t\), inferred from the second image, from the appearance of the first distorted image.}
    \item \highlight{The appearance of the first distorted image is predicted by adding the effect of the transform \(t\), inferred from the second image, to the appearance of the first pristine image.}
\end{itemize}
\highlight{The other two transforms, corresponding to the second image pair, may be interpreted similarly.}

To obtain crossed reconstructions, we apply \textbf{content shuffling}, \highlight{as in prior style-transfer methods \cite{ref:drit} \cite{ref:disrep_domain},} to replace content features across domains. \highlight{Hence, the crossed reconstruction predictions are obtained using the decoder:}
\begin{align}
    \highlight{\Tilde{y}_{11} = D_\theta\left(c_{12}, \Tilde{a}_{11}\right)}, \nonumber \\
    \highlight{\Tilde{y}_{12} = D_\theta\left(c_{11}, \Tilde{a}_{12}\right)}, \nonumber \\
    \highlight{\Tilde{y}_{21} = D_\theta\left(c_{22}, \Tilde{a}_{21}\right)}, \nonumber \\
    \highlight{\Tilde{y}_{22} = D_\theta\left(c_{21}, \Tilde{a}_{22}\right)}.
\end{align}
\highlight{The content features used for cross-reconstructions are denoted by \(\Tilde{c}_{ij}\) in Fig. \ref{fig:disque_training} for notational convenience. For example, \(\Tilde{c}_{11} = c_{12}\).}

\highlight{The first two cross-reconstructions may be interpreted as follows:}
\begin{itemize}
    \item \highlight{The first pristine image is inferred by combining the content feature of the first distorted image with the predicted appearance feature of the first pristine image.}
    \item \highlight{The first distorted image is inferred by combining the content feature of the first pristine image with the predicted appearance feature of the first distorted image.}
\end{itemize}
\highlight{The other two cross-reconstructions follow a similar interpretation.}

These crossed reconstructions are evaluated using the cross-reconstruction objective
\begin{equation}
    \mathcal{L}_{cross} = \sum\limits_{ij} \mathcal{L}_{char}(x_{ij}, \Tilde{y}_{ij}) + \lambda_f \mathcal{L}_{freq}(x_{ij}, \Tilde{y}_{ij}).
\end{equation}
The effect of using appearance mixing to mitigate the effect of CAFs is illustrated in Section \ref{sec:evaluation_tonemapping}.

Since \(x_{i*}\) are views of the same image, we expect content representations generated by a good content encoder to satisfy
\begin{equation}
    c_{11} \approx c_{12}, \quad c_{21} \approx c_{22}.
\end{equation}
Moreover, since the two input images were transformed by the same transformation \(t\), we expect appearance representations generated by a good appearance encoder to satisfy
\begin{equation}
    \Delta a_1 \approx \Delta a_2.
\end{equation}

We guide the networks to learn these properties using a symmetrized InfoNCE \cite{ref:infonce} contrastive loss, as in MoCov3 \cite{ref:mocov3}. InfoNCE loss aims to maximize the similarity between a query-positive key pair \(\left(q, k^{+}\right)\), \highlight{obtained from the same data sample,} while minimizing the similarity between the query and a set of negative keys \highlight{\(\{k^{-}\}\), which are obtained from other data samples}:
\begin{equation}
    \highlight{\mathcal{L}_\mathit{InfoNCE} = -\log\frac{\exp\left(q \cdot k^{+} / \tau\right)}{\exp\left(q \cdot k^{+} / \tau\right) + \sum\limits_{\{k^{-}\}} \exp\left(q \cdot k^{-} / \tau\right)}.}
\end{equation}

Given a batch of training samples, the ``content contrastive loss'' \(\mathcal{L}_{c, \mathit{InfoNCE}}\) is obtained by using \(\left([c_{11}, c_{21}], [c_{12}, c_{22}]\right)\) from the same sample as positive query-key pairs and those from different samples in the batch as negative pairs. Similarly, the ``appearance contrastive loss'' \(\mathcal{L}_{a, \mathit{InfoNCE}}\) is obtained by using \(\left(\Delta a_1, \Delta a_2\right)\) from the same sample as positive query-key pairs and those from different samples as negative pairs.

\begin{equation}
\mathcal{L}_{c, \mathit{InfoNCE}} \sim \left([c_{11}, c_{21}], [c_{12}, c_{22}]\right)
\end{equation}

\begin{equation}
\mathcal{L}_{a, \mathit{InfoNCE}} \sim \left(\Delta a_1, \Delta a_2\right)
\end{equation}
Hence, the parameters of \(\left(C_\theta, A_\theta, D_\theta\right)\) are trained to minimize the overall learning objective
\begin{equation}
    \mathcal{L} = \left(\mathcal{L}_{self} + \mathcal{L}_{cross}\right) + \beta \left(\mathcal{L}_{c, \mathit{InfoNCE}} + \mathcal{L}_{a, \mathit{InfoNCE}}\right).
\end{equation}
A visualization of computation of the training objective is depicted in Fig. \ref{fig:disque_training}.

%% file: sections/disque_subsections/architecture.tex
\subsection{Network Architecture}
\label{sec:disque_architecture}

\highlight{To enable a fair comparison against prior work on self-supervised representation learning for quality assessment} \cite{ref:contrique} \cite{ref:reiqa}, we adopted a ResNet-50-based architecture for both the content and appearance encoders, and a reversed ResNet-50 architecture for the decoder, with key modifications. Due to the presence of two encoders and one decoder, we term this architecture a ``dual-head'' U-Net \cite{ref:unet}. To limit the number of features used in downstream picture quality assessment tasks, we adopted a ResNet-50 0.5x architecture, i.e., one that uses half the number of channels at each layer. The other departure from ResNet-50 was the removal of batch normalization layers, since they have been shown to hinder image-to-image translation performance \highlight{by reducing the range of intermediate feature activations} \cite{ref:enhanced_resnet} \cite{ref:dynamic_deblur}. In particular, removing batch normalization has been shown to improve super-resolution accuracy in \cite{ref:enhanced_resnet}.

Moreover, we introduced instance normalization (IN) \cite{ref:instance_norm} layers to ResNet blocks in the content encoder to introduce the effect of ``appearance normalization.'' IN layers normalize the statistics of each channel of the input feature map. Consider a feature map \(F_{ncij} \in \mathbb{R}^{N\times C \times H \times W}\). Then, the output of the IN layer is
\begin{equation}
    \Tilde{F}_{nchw} = \frac{F_{nchw} - \mu_{nc}}{\sigma_{nc}},
\end{equation}
where
\begin{equation}
    \mu_{nc} = \frac{1}{HW} \sum\limits_{ij} F_{ncij}
\end{equation}
and
\begin{equation}
    \sigma_{nc} = \sqrt{\frac{1}{HW} \sum\limits_{ij} \left(F_{ncij} - \mu_{nc}\right)^2}.
\end{equation}

This follows prior work in style transfer that uses IN layers for style transfer \cite{ref:instance_norm} \cite{ref:condin}, which demonstrated that mean and standard deviations of layer activations may be used to encode ``style.'' As a result, normalizing these statistics using IN layers was found to improve style transfer performance. Hence, we deploy IN layers to normalize appearance and retain only content-related features.

By contrast, the appearance encoder does not include IN layers since its goal is to capture appearance information. This is achieved by average pooling intermediate layer feature activations obtained from each ResNet block. Therefore, the appearance of the input to the network is captured by a single feature vector rather than a spatially-varying feature map. Despite being assumed to be slow-varying over space, appearance is a non-stationary attribute of images. For example, one region of an image may have bright objects while another has dark objects. Hence, the dual-head U-Net is best applied on small image patches, rather than on full images. Here, we use a patch size of 128\(\times\)128.

\begin{figure*}
    \centering
    \includegraphics[width=0.75\linewidth]{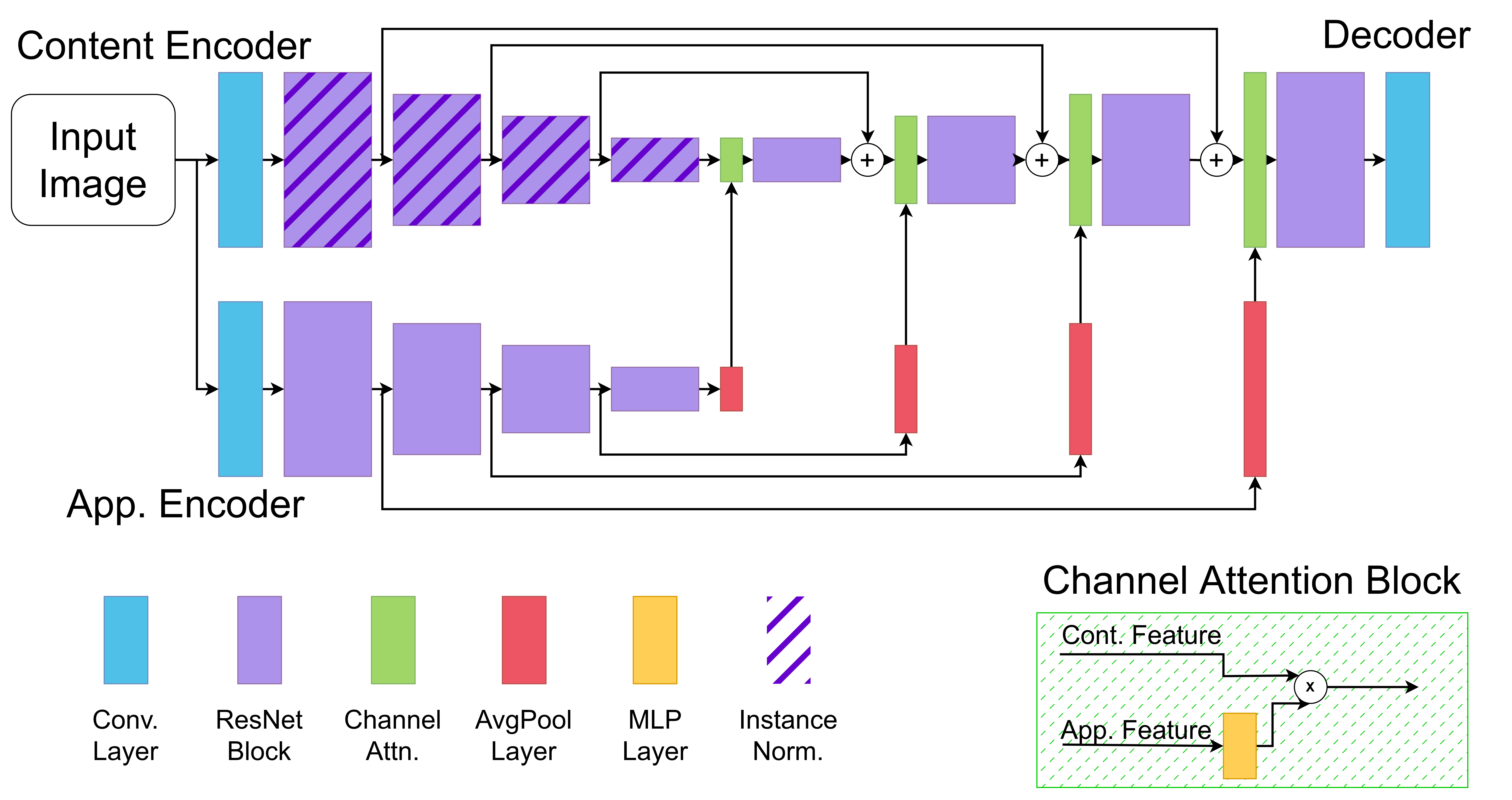}
    \caption{The dual-head U-Net architecture.}
    \label{fig:architecture}
\end{figure*}

The decoder follows a typical U-Net structure, using skip connections to introduce multi-level feature maps from the content encoder. Appearance features are introduced into the encoder using a product-based channel attention mechanism \(CA(x, a) = x \otimes a\), similar to that used in residual attention networks \cite{ref:ranet}. We chose this mechanism since channel attention may be considered an inverse of instance normalization that re-introduces the desired appearance features, as evidenced by Adaptive Instance Normalization \cite{ref:adain}. The overall dual-head U-Net structure of the proposed deep network is illustrated in Fig. \ref{fig:architecture}.

%% file: sections/disque_subsections/quality_assessment.tex
\subsection{Visual Quality Assessment}
\label{sec:disque_quality_assessment}
After training, we use the appearance encoder to conduct FR visual quality assessment. Because the encoder was trained to disentangle appearance information from image content, we term our quality predictor the \textbf{Disentangled Quality Evaluator (DisQUE)}.

Given a pair of reference and test images \(I_{ref}\) and \(I_{dis}\), we obtain feature maps from the output of each of the four ResNet blocks in \(A_\theta\):
\begin{equation}
    \mathcal{A}_{ref} = A_\theta(I_{ref}), \quad \mathcal{A}_{dis} = A_\theta(I_{dis}).
\end{equation}
We then characterized feature maps by computing both the mean and standard deviation of each channel
\begin{equation}
    a_{\mu} = E\left[\mathcal{A}\right], \quad a_{\sigma} = \sqrt{E\left[\left(\mathcal{A} - a_{\mu}\right)^2\right]}.
\end{equation}

Although computing the mean is typical, standard deviations of feature maps have also been used \cite{ref:msml}, albeit without explicit justification. Here, we justify the use of standard deviation by referring to one of the assumptions behind our disentangled representation model. We posited that appearance varies slowly over space, because of which we used spatially constant appearance vectors to describe image patches. However, since quality assessment is carried out over images, the standard deviation captures variations in appearance over space. Hand-crafted quality models such as ESSIM \cite{ref:essim} and GMSD \cite{ref:gmsd} have also benefited from characterizing spatial quality variations using standard deviations.

Following prior work \cite{ref:contrique} \cite{ref:reiqa} \cite{ref:msml}, we captured multi-scale appearance features by repeating the process using input images rescaled to half resolution. The final feature vector for the reference and test images \(\left(z_{ref}, z_{dis}\right)\) was obtained by concatenating mean and standard deviation-pooled features at both scales. Since DisQUE is an FR quality model, the difference in features between the two images \(z = \left|z_{ref} - z_{dis}\right|\) was used to predict quality.

The use of multi-scale, multi-block features pooled spatially using two methods yields a feature vector of size 8192. Finally, a linear regressor model was used to map the appearance features to subjective quality scores. Note that the self-supervised appearance network was frozen during inference time, and only the linear regressor was recalibrated on each evaluation dataset. Ablation experiments studying the effects of multi-scale features and the use of standard deviation pooling are presented in Section \ref{sec:evaluation}.

%% file: sections/datasets.tex
\section{Datasets}
\label{sec:datasets}

\input{sections/datasets_subsections/hdr}
\input{sections/datasets_subsections/sdr}

%% file: sections/datasets_subsections/hdr.tex
\subsection{HDR Datasets}
\label{sec:datasets_hdr}
To train DisQUE for HDR quality assessment, we used the recently developed LIVE UGC-HDR database \cite{ref:live_ugc_hdr}, which is the first publicly available large-scale database of HDR videos. The database consists of over 2,153 HLG-encoded \cite{ref:hlg} videos filmed by amateur iPhone users, containing a diverse collection of scenes, including indoor, outdoor, daytime, nighttime, static, and dynamic scenes containing camera and object motion.

Since the proposed DRL method learns from images, we first sampled video frames from the set of HDR videos. To introduce sufficient content diversity, we sampled video frames at 2-second intervals, yielding a total of 19060 frames at an average of 8.85 frames per video. Since the dataset contains both 1080p and 4K videos, we rescaled all videos to 1080p using Lanczos rescaling. Finally, since the PQ \cite{ref:pq} standard can represent a wider range of brightnesses, we re-encoded all sampled frames to 10-bit PQ from 10-bit HLG.

As described earlier, we aimed to train DisQUE to predict the quality of tone-mapped and compressed HDR videos. So, we used a bank of transforms (called \(\mathcal{T}\) in Section \ref{sec:disque_learning}) consisting of the following ten open-source tone-mapping operators (TMOs), with their parameters varied to generate a diverse set of tone mapping-related distortions.

\begin{itemize}
\item \textbf{Hable} \cite{ref:hable} - A parameter-free pointwise non-linear transform originally designed for use in the video game \textit{Uncharted 2}. A desaturation parameter was varied to control how colorful tone-mapped images would appear.

\item \textbf{Reinhard02} \cite{ref:reinhard02} - A point non-linearity to map luminances from HDR to SDR. A desaturation parameter was varied to control how colorful tone-mapped images would appear.

\item \textbf{Durand02} \cite{ref:durand02} - Uses a ``fast bilateral filter'' to decompose the luminances of HDR frames into ``base'' and ``detail'' layers. A contrast parameter was varied to control the degree of global contrast, i.e., the difference between the visibilities of bright and dark regions.

\item \textbf{Shan12} \cite{ref:shan12} - Uses an edge-aware stationary wavelet transform (SWT) \cite{ref:swt}. The number of wavelet levels was varied, which affected contrast.

\item \textbf{Reinhard12} \cite{ref:reinhard12} - Uses color-appearance models applied in a local manner. The assumed viewing conditions were varied, which introduced color distortions.

\item \textbf{Eilertsen15} \cite{ref:eilertsen15} - Applies a ``fast detail extraction'' method to obtain a base-detail decomposition and applies a dynamic tone-curve. The coarseness of the tone curve was varied, which led to contrast distortions.

\item \textbf{Oskarsson17} \cite{ref:oskarsson17} - Uses Dynamic Programming to cluster values in the input image channels. The number of clusters was varied, which introduced quantization artifacts such as banding.

\item \textbf{Rana19} \cite{ref:rana19} - Uses a Generative Adversarial Network (GAN) to create a fully-convolutional, parameter-free TMO. A desaturation parameter was varied to control how colorful tone-mapped images appeared.

\item \textbf{Yang21} \cite{ref:yang21} - Uses a deep convolutional neural network (CNN) to transform a multi-scale Laplacian pyramid decomposition of each input HDR frame. A desaturation parameter was varied to control how colorful tone-mapped images appeared.

\item \textbf{ITU21} \cite{ref:itu21} - A parameter-free TMO proposed by the ITU in Recommendation BT.2446 (``Approach A''). The nominal HDR luminance was varied, which affected the brightness and contrast of tone-mapped images.
\end{itemize}

Furthermore, we introduced compression distortions by applying lossy JPEG compression at four levels to the tone-mapped images. Therefore, each transform in the bank \(\mathcal{T}\) consists of tone-mapping using one of the aforementioned TMOs followed by JPEG compression.

We tested the efficacy of DisQUE on the LIVE Tone-Mapped HDR (LIVE-TMHDR) subjective database \cite{ref:live_tmhdr}, which is the first public database of subjectively annotated tone-mapped and compressed HDR videos. LIVE-TMHDR consists of 15,000 distorted videos that were generated from 40 source contents (20 each encoded using PQ and HLG) using 13 tone-mapping methods and compressed using libx264 \cite{ref:x264} at three quality levels. The 13 tone-mapping methods include the 10 TMOs discussed here, Dolby Vision tone-mapping \cite{ref:dolbyvision}, the Color Space Transform (CST) method used for gamut/tone mapping by colorists, and manual tone-mapping by a human expert colorist. Moreover, the TMOs were applied to videos using three ``temporal modes,'' which varied the degree of temporal distortions.

%% file: sections/datasets_subsections/sdr.tex
\subsection{SDR Datasets}
\label{sec:datasets_sdr}
To demonstrate the versatility of DisQUE, we also evaluated its performance on SDR FR quality assessment. We followed a similar approach as prior work \cite{ref:reiqa}, \cite{ref:contrique} to create a training dataset of SDR images from the following diverse sources\footnote{The SDR training dataset was collected, processed, and run at the University of Texas at Austin by university-affiliated authors.}.
\begin{itemize}
    \item KADIS-700k \cite{ref:kadis700k} - \(\sim\)140K images
    \item AVA \cite{ref:ava} - \(\sim\) 255K images
    \item CERTH-Blur \cite{ref:blur} - \(\sim\) 2.5K images
    \item VOC \cite{ref:voc} - \(\sim\) 33K images
    \item COCO \cite{ref:coco} - \(\sim\) 330K images
    \item Places \cite{ref:places} - \(\sim\) 2.2M images
\end{itemize}
In total, we obtained nearly 3M images from these data resources. As we will describe below, the bank of transforms used for SDR training includes color and contrast distortions. So, we excluded grayscale images and those having significant over/under-exposed regions to create a training dataset of nearly 1.8M training images.

The bank of transforms for SDR training was constructed using the following set of 25 distortions borrowed from \cite{ref:reiqa}, which may be applied at five degrees of severity each. \highlight{The set of distortions includes various types of blur, noise, resizing, and color distortions. A full list is provided in the supplementary material.}

Each of the aforementioned methods typically modifies only one aspect of each image, such as color, brightness, etc. However, the real world may present complex combinations of distortions. To simulate these scenarios, we constructed each transform in \(\mathcal{T}\) as a composition of one to three randomly chosen unit distortions, each applied at a randomly chosen level of severity.

We evaluated the SDR DisQUE model on the four FR picture quality assessment datasets. The LIVE-IQA dataset \cite{ref:live_iqa} consists of 29 reference images subjected to five distortions - blur, noise, JPEG compression, JPEG2000 compression, and bit errors in JPEG2000 bitstreams. This procedure yielded a total of 982 distorted pictures. The CSIQ dataset \cite{ref:csiq} consists of 866 test images generated from 30 source contents subjected to six distortions - blur, noise, JPEG compression, JPEG2000 compression, pink Gaussian noise, and global contrast decrements.

TID2013 \cite{ref:tid2013} is a dataset of 3000 test images generated by applying 24 impairments at five levels each to a dataset of 25 images. The set of distortions includes blur, noise, compression, bitstream errors, contrast and color distortions, and spatial distortions such as jitter and color blocking. Finally, KADID-10k \cite{ref:kadid10k} is the largest database on the list, containing 10,125 test images generated by subjecting 81 pristine images to 25 distortions at five levels each. The types of distortions in KADID-10k are similar to those in TID2013.

%% file: sections/evaluation.tex
\section{Evaluation}
\label{sec:evaluation}

\input{sections/evaluation_subsections/training}
\input{sections/evaluation_subsections/quality}
\input{sections/evaluation_subsections/tonemapping}

%% file: sections/evaluation_subsections/training.tex
\subsection{Training}
\label{sec:evaluation_training}
In both the HDR and SDR cases, the dual-head U-Net models were trained using a batch size of 36, split across 9 NVIDIA A-100 GPUs. Note that each sample consists of four randomly sampled and transformed (using \(t \sim \mathcal{T}\)) \(128 \times 128\) image patches \(\left(x_{11}, x_{12}, x_{21}, x_{22}\right)\), as described in Section \ref{sec:disque_learning}. The dual-head U-Net was trained for 400K steps using an Adam optimizer configured with an initial learning rate of 0.0002. The learning rate was decayed by 0.99 every 1,000 steps, and the loss hyperparameters were set to \(\lambda_f = 0.1\) and \(\beta = 0.5\).

%% file: sections/evaluation_subsections/quality.tex
\subsection{Evaluating HDR and SDR Quality Prediction}
\label{sec:evaluation_quality}

\begin{table}[t]
    \centering
    \caption{Evaluation of Quality Prediction Models on LIVE-TMHDR}
    \label{tab:live_tmhdr_accs}
    \begin{tabular}{|c|c|c|c|c|c|c|c|}
        \hline
         Type & Model & PCC & SROCC & RMSE \\
         \hline \\[-9pt]
        \multirow{13}{*}{Hand-crafted} & Y-FUNQUE+ & 0.4524 & 0.4343 & 9.4352 \\
        & BTMQI & 0.4705 & 0.4663 & 9.2238 \\
        & FSITM & 0.4813 & 0.4626 & 8.9212 \\
         & NIQE & 0.4805 & 0.4746 & 9.5563 \\
         & BRISQUE & 0.4811 & 0.4833 & 8.9869 \\
         & DIIVINE & 0.4794 & 0.4925 & 9.2879 \\
         & TMQI & 0.5062 & 0.4956 & 8.6897 \\
         & FUNQUE & 0.5082 & 0.4949 & 8.8863 \\
         & TMVQI & 0.5198 & 0.4969 & 8.8697 \\
         & FFTMI & 0.5298 & 0.5315 & 8.8559 \\
         & 3C-FUNQUE+ & 0.5817 & 0.5661 & 8.6568 \\
         & HIGRADE & 0.6682 & 0.6698 & 8.2619 \\
         & Cut-FUNQUE & 0.7783 & 0.7781 & 6.4187 \\
        \hline
        \multirow{5}{*}{Deep Networks} & RcNet & 0.5985 & 0.5824 & 8.2417 \\
         & CONTRIQUE &  0.7360 & 0.7230 & 6.8476 \\
         & ReIQA & 0.7583 & 0.7812 & 7.2951 \\
         & MSML & 0.7883 & 0.7740 & 6.8090 \\
        & \textbf{DisQUE} & \textbf{0.8160} & \textbf{0.8215} & \textbf{6.3241} \\
         \hline
    \end{tabular}
\end{table}

\begin{table*}[t]
    \centering
    \caption{Evaluation of Quality Prediction Models on SDR Quality Databases}
    \label{tab:sdr_accs}
    \begin{tabular}{|c|c|c|c|c|c|c|c|c|c|}
        \hline
         \multirow{2}{*}{Type} & \multirow{2}{*}{Model} & \multicolumn{2}{c|}{LIVE IQA} & \multicolumn{2}{c|}{CSIQ} & \multicolumn{2}{c|}{TID2013} & \multicolumn{2}{c|}{KADID-10k} \\
         \cline{3-10} & \\[-8pt]
         & & PCC & SROCC & PCC & SROCC & PCC & SROCC & PCC & SROCC \\
         \hline
         \multirow{5}{*}{Hand-crafted} & PSNR & 0.868 & 0.881 & 0.824 & 0.820 & 0.675 & 0.643 & 0.680 & 0.677 \\
         & BRISQUE \cite{ref:brisque} & 0.935 & 0.939 & 0.829 & 0.746 & 0.694 & 0.604 & 0.567 & 0.528 \\
         & SSIM \cite{ref:ssim} & 0.911 & 0.921 & 0.835 & 0.854 & 0.698 & 0.642 & 0.633 & 0.641 \\
         & FSIM \cite{ref:fsim} & 0.954 & 0.964 & 0.919 & 0.934 & 0.875 & 0.852 & 0.850 & 0.854 \\
         & CORNIA \cite{ref:cornia} & 0.950 & 0.947 & 0.776 & 0.678 & 0.768 & 0.678 & 0.558 & 0.516 \\
         \hline
         \multirow{4}{*}{Supervised Deep Nets} & DB-CNN \cite{ref:dbcnn} & 0.971 & 0.968 & 0.959 & 0.946 & 0.865 & 0.816 & 0.856 & 0.851 \\
         &  PQR \cite{ref:pqr} & 0.971 & 0.965 & 0.901 & 0.872 & 0.798 & 0.740 & - & - \\
         & HyperIQA \cite{ref:hyperiqa} & 0.966 & 0.962 & 0.942 & 0.923 & 0.858 & 0.840 & 0.845 & 0.852 \\
         & LoDA \cite{ref:loda} & 0.979 & 0.975 & 0.901 & 0.869 & - & - & 0.936 & 0.931 \\
         & DRF-IQA \cite{ref:drf_iqa} & 0.983 & 0.983 & 0.960 & 0.964 & 0.942 & 0.944 & - & - \\
         & RKIQT \cite{ref:rkiqt} & 0.986 & 0.984 & 0.970 & 0.958 & 0.917 & 0.900 & 0.911 & 0.911 \\
         & SaTQA \cite{ref:satqa} & 0.983 & 0.983 & 0.972 & 0.965 & 0.948 & 0.938 & 0.949 & 0.946 \\
         & DRIQA \cite{ref:driqa} & 0.989 & 0.985 & 0.980 & 0.978 & 0.961 & 0.954 & - & - \\
         \hline
         \multirow{4}{*}{Self-Supervised Deep Nets} & LPIPS \cite{ref:lpips} & 0.936 & 0.932 & 0.906 & 0.884 & 0.756 & 0.673 & 0.713 & 0.721 \\
         & CONTRIQUE \cite{ref:contrique} & 0.966 & 0.966 & \textbf{0.964} & 0.956 & \textbf{0.915} & \textbf{0.909} & \textbf{0.947} & \textbf{0.946} \\
         & ReIQA \cite{ref:reiqa} & \textbf{0.974} & \textbf{0.973} & \textbf{0.965} & \textbf{0.961} & \textbf{0.915} & 0.905 & 0.903 & 0.901 \\
         & \textbf{DisQUE} & \textbf{0.972} & \textbf{0.970} & 0.956 & \textbf{0.961} & 0.909 & \textbf{0.922} & \textbf{0.921} & \textbf{0.934} \\
         \hline
    \end{tabular}
\end{table*}

\begin{table*}[t]
    \centering
    \caption{Ablation Experiments Studying the Effect of Multi-Scale and Multi-Pooling Features}
    \label{tab:scale_pool_ablation}
    \begin{tabular}{|c|c|c|c|c|c|c|c|c|c|c|}
        \hline
         \multirow{2}{*}{DisQUE Variant} & \multicolumn{2}{c|}{LIVE-TMHDR} & \multicolumn{2}{c|}{LIVE IQA} & \multicolumn{2}{c|}{CSIQ} & \multicolumn{2}{c|}{TID2013} & \multicolumn{2}{c|}{KADID-10k} \\
         \cline{2-11} & \\[-8pt]
         & PCC & SROCC & PCC & SROCC & PCC & SROCC & PCC & SROCC & PCC & SROCC \\
         \hline
         Single-Scale, Mean Pooling & 0.767 & 0.762 & 0.961 & 0.956 & 0.926 & 0.940 & 0.865 & 0.879 & 0.889 & 0.907 \\
         Multi-Scale, Mean Pooling & 0.803 & 0.807 & 0.968 & 0.961 & 0.947 & 0.951 & 0.891 & 0.898 & 0.908 & 0.923 \\
         Single-Scale, Mean+Std Pooling & 0.805 & 0.804 & 0.967 & 0.964 & 0.948 & 0.956 & 0.900 & 0.915 & 0.910 & 0.928 \\
         \textbf{Multi-Scale, Mean+Std Pooling} & \textbf{0.816} & \textbf{0.822} & \textbf{0.972} & \textbf{0.970} & \textbf{0.956} & \textbf{0.961} & \textbf{0.909} & \textbf{0.922} & \textbf{0.921} & \textbf{0.934}\\
         \hline
    \end{tabular}
\end{table*}
We evaluated DisQUE's HDR tone-mapping quality predictions on the LIVE-TMHDR video quality dataset, and SDR quality prediction on four datasets - LIVE IQA, CSIQ, TID2013, and KADID-10k. In all cases, as described in Section \ref{sec:disque_quality_assessment}, DisQUE generated an 8192-dimensional feature vector by applying the appearance encoder to both the reference and test pictures/video frames. For video quality prediction on LIVE-TMHDR, we averaged the feature vector obtained from each frame to yield a video-level feature vector.

We evaluated quality prediction accuracy on each dataset using a 10-fold random cross-validation, where 80\% of the dataset was used to train the PLS projector and the Linear SVR predictor, and 20\% was used for testing. The hyperparameters of the Linear SVR predictor were selected by performing five-fold cross-validation on the training dataset of each random cross-validation split. In addition to the combination of PLS and Linear SVR, we also experimented with Lasso and Ridge regressors, and the best regression model was chosen. Note that all three approaches yield linear prediction models and so, incur similar computational complexities at inference time. The best regressor models and their hyperparameters were identified by optimizing the average of the Pearson Correlation Coefficient (PCC) and Spearman Rank Order Correlation (SROCC) between the predicted and ground-truth subjective scores on the validation datasets.

The quality prediction accuracies of various hand-crafted and deep quality models on the LIVE-TMHDR database are presented in Table \ref{tab:live_tmhdr_accs}, while the SDR quality prediction outcomes are presented in Table \ref{tab:sdr_accs}. From the Tables, it may be observed that DisQUE outperformed all the compared models compared in Table \ref{tab:live_tmhdr_accs} on the tone-mapping quality prediction task, and achieved comparable state-of-the-art (SOTA) accuracy among the compared self-supervised models on the SDR quality prediction task. \highlight{The variation in quality prediction accuracy between datasets is generally explained by the sizes and variety of distortions present in the four datasets. While LIVE-IQA and CSIQ databases are small and contain a small set of distortions, TID2013 and KADID-10k datasets are larger databases with a wider variety of distortions. Moreover, some distortions in TID2013 and KADID-10k, such as masked noise, speckle noise, and jitter,  may be unfamiliar to average human subjects. This may lead to higher variability in subject ratings, making quality prediction more difficult. Due to these reasons, most quality prediction algorithms tend to achieve lower quality prediction accuracies on TID2013 and KADID-10k compared to LIVE-IQA and CSIQ.}

We further analyzed the effect of feature subsets of DisQUE on quality prediction accuracy in an ablation study. As explained in Section \ref{sec:disque_quality_assessment}, DisQUE combines four subsets of features generated by applying mean and standard deviation pooling to the appearance encoder's feature maps. The impact of introducing each feature subset into DisQUE is quantified in Table \ref{tab:scale_pool_ablation}. From this Table, it may be seen that both multi-scale features and standard-deviation pooling improved quality prediction accuracy across databases.

%% file: sections/evaluation_subsections/tonemapping.tex
\subsection{Example-Guided Tone Mapping}
\label{sec:evaluation_tonemapping}
\begin{figure*}
    \centering
    \subfloat[{Example-guided contrast tuning}\label{fig:distm_contrast}]{%
      \includegraphics[width=0.45\linewidth]{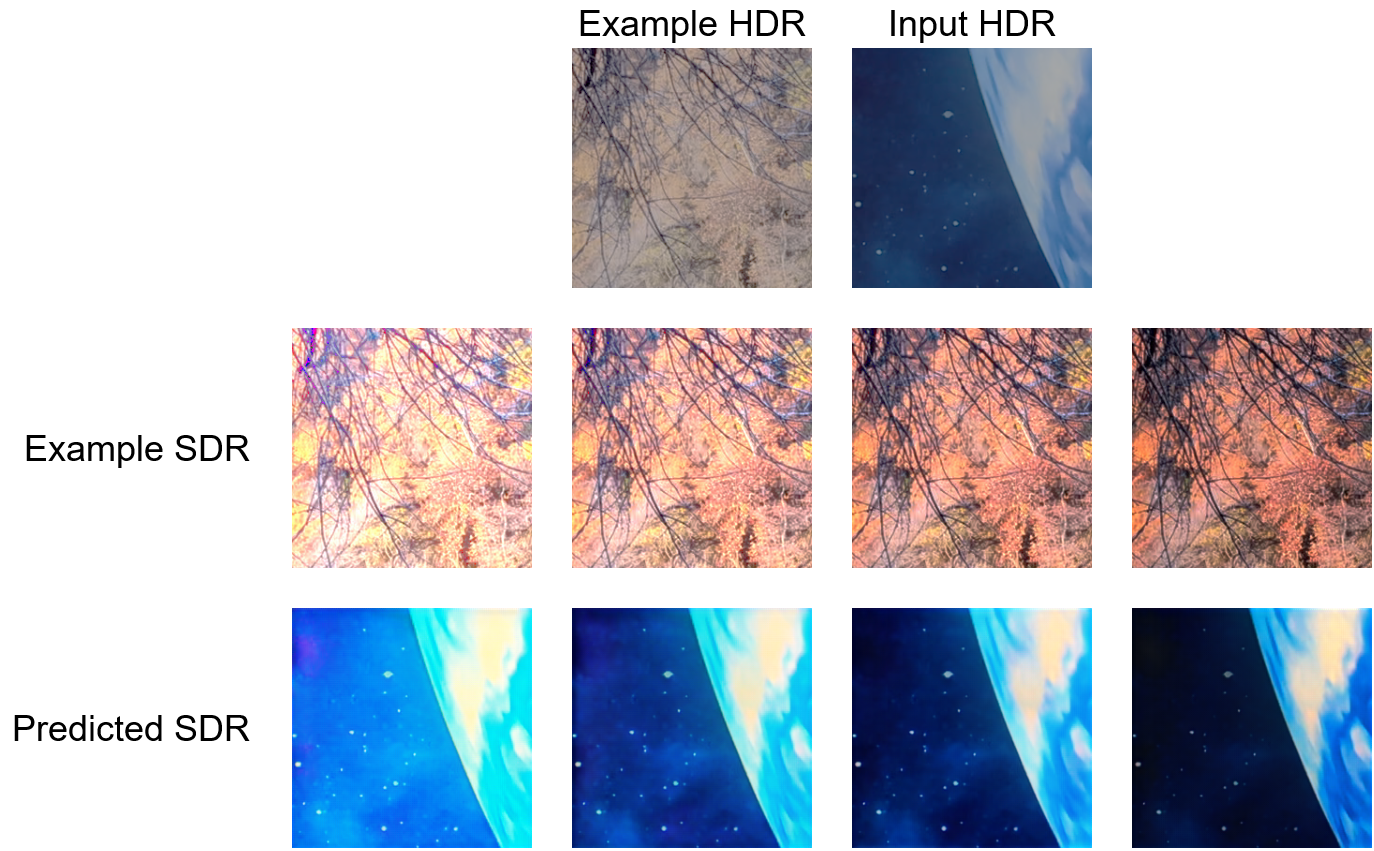}
    }%
    \subfloat[{Example-guided color desaturation tuning}\label{fig:distm_desat}]{%
      \includegraphics[width=0.45\linewidth]{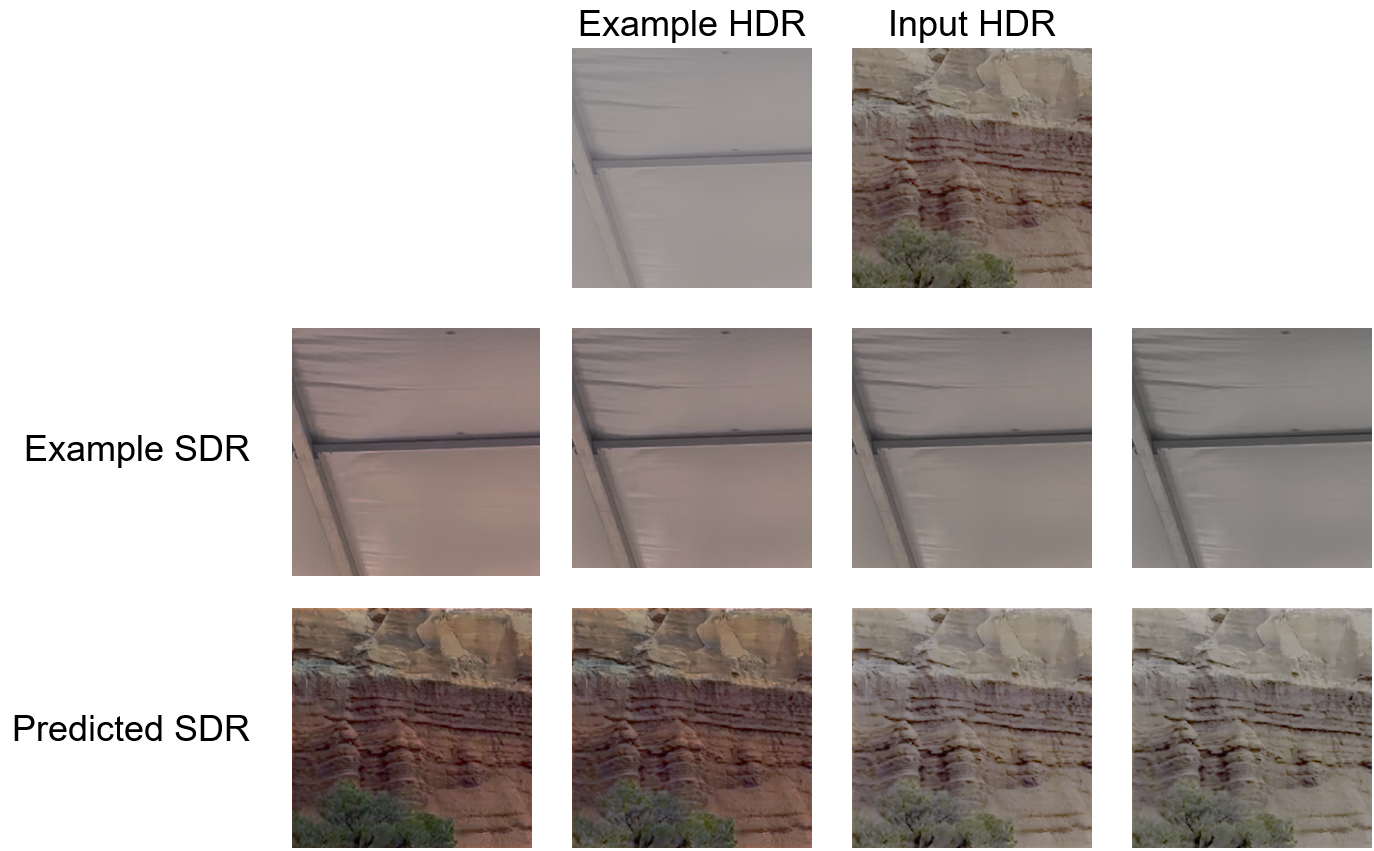}
    }
    \\
    \subfloat[{Example-guided brightness tuning}\label{fig:distm_brightness}]{%
      \includegraphics[width=0.45\linewidth]{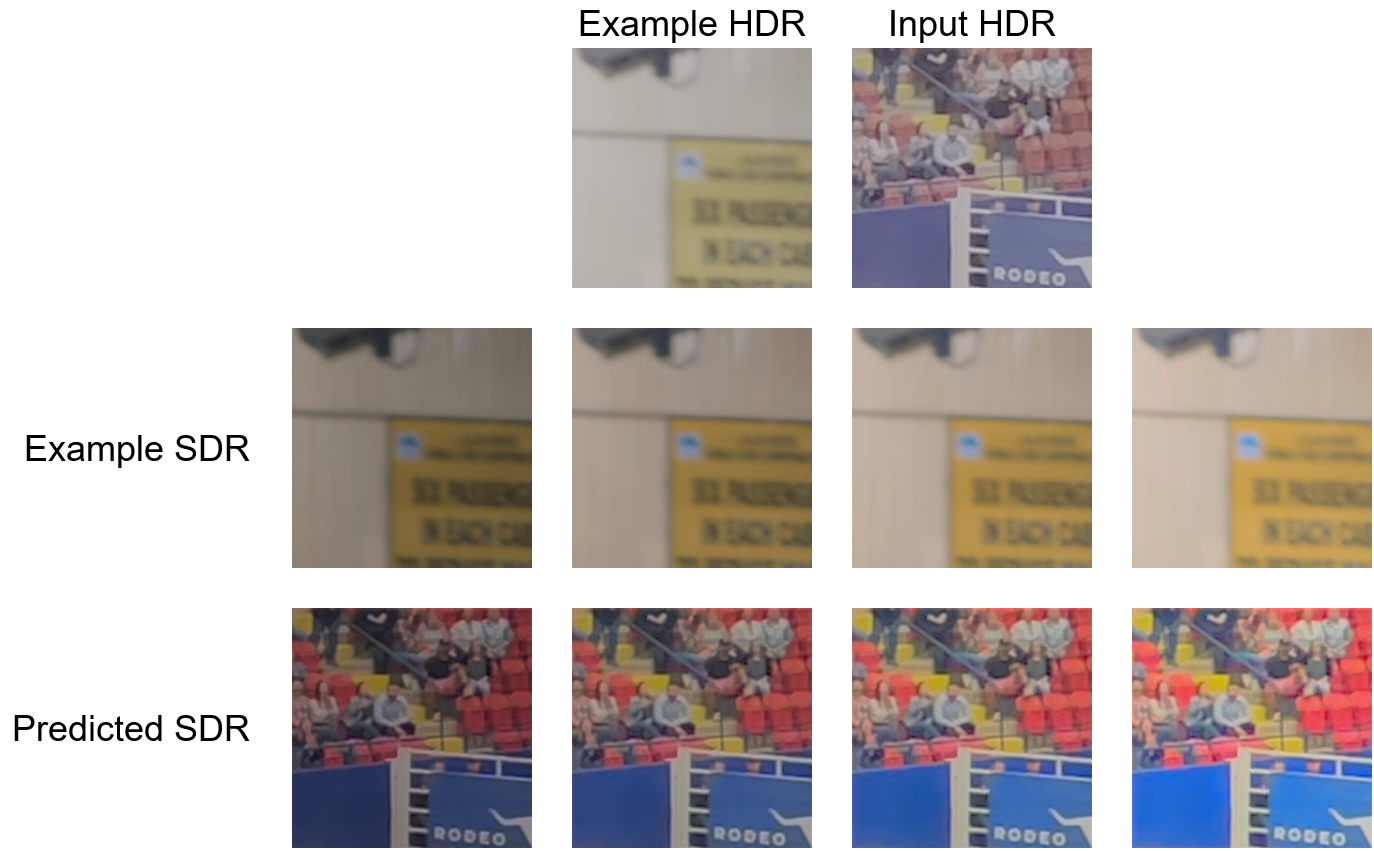}
    }%
    \subfloat[{Example-guided global hue tuning}\label{fig:distm_hue}]{%
      \includegraphics[width=0.45\linewidth]{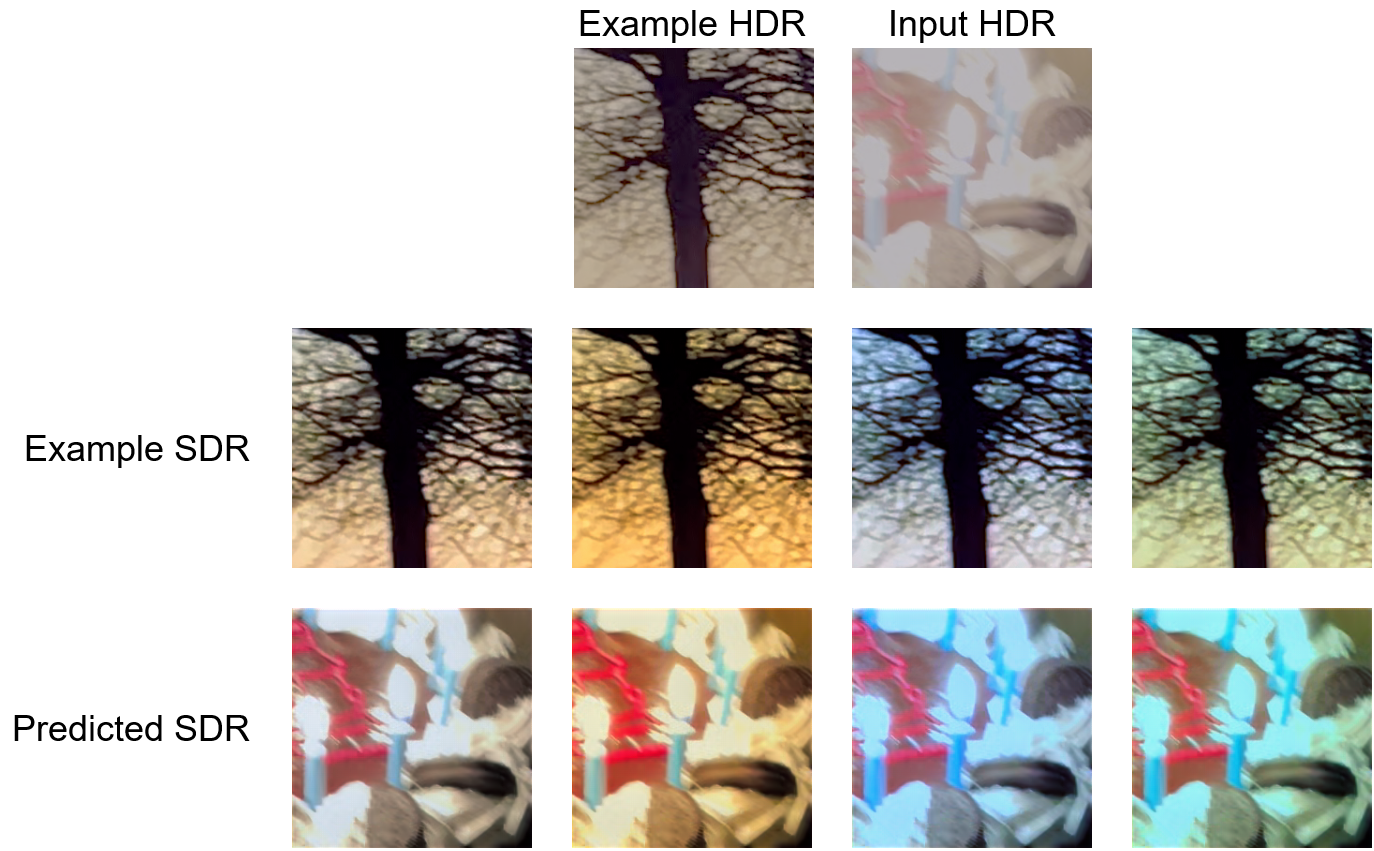}
    }
    \caption{Examples of using example-driven tone mapping to vary tone-mapping characteristics}
    \label{fig:distm_samples}
\end{figure*}

\begin{figure}
    \centering
    \includegraphics[width=0.8\linewidth]{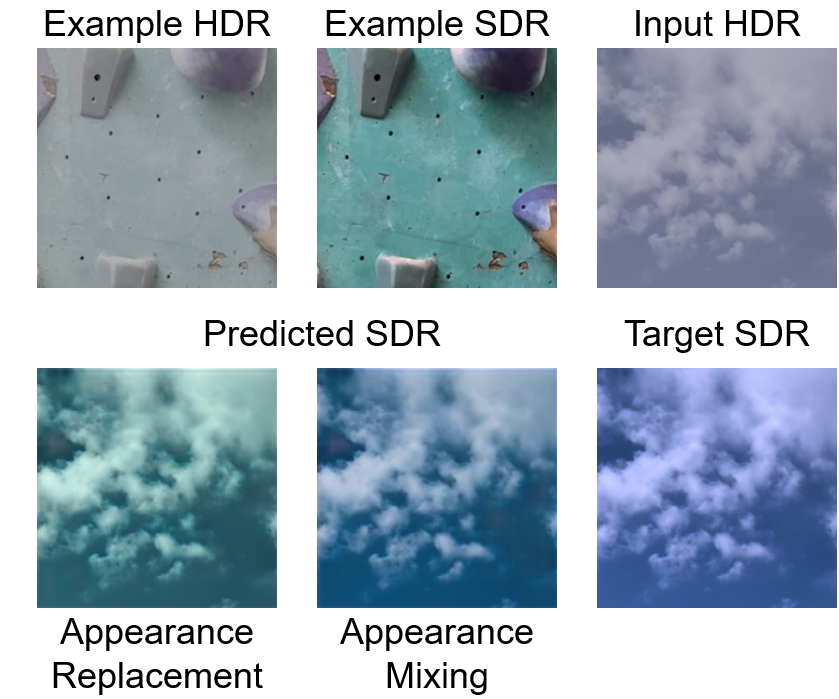}
    \caption{Effect of confounding appearance features on EGTM.}
    \label{fig:distm_confounding}
\end{figure}

A key element of training DisQUE is the use of appearance mixing to yield crossed predictions \(\Tilde{y}_{ij}\). In addition to being a representation learning framework, the DisQUE training paradigm may also be used to perform \textbf{example-guided image processing (EGIP)}.

We define an EGIP task as consisting of three inputs - the example source, the example target, and the input source image. The EGIP model then processes the input source image to induce a similar effect as shown by the example pair. For example, if the example target is a blurred version of the example source, an ideal EGIP network predicts the target image as a blurred version of the input source. EGIP, as defined here, may be seen as a task this is complementary to guided image filtering \cite{ref:guided_filtering}, where content from a guidance image (analogous to the example pair here) is transferred to the input, rather than appearance.

Here, we demonstrate the EGIP capability of the disentangling network by performing example-guided HDR tone mapping (EGTM). Fig. \ref{fig:distm_samples} shows examples of using EGTM to vary visual characteristics such as color, contrast, and brightness of tone-mapped images. We achieved this by varying the corresponding attribute of the example SDR image, and it may be seen that the network was able to characterize the differences between the example HDR and SDR images and transfer those characteristics to the input HDR image to predict the corresponding tone-mapped SDR image. \highlight{It is important to note that in all experiments shown here, the example image was chosen at random, independent of the target image. Future works may explore methods to choose examples, say from a database, that are similar in visual characteristics to the input HDR image.}

The EGTM results presented in Fig. \ref{fig:distm_samples} all use the appearance mixing method described in Section \ref{sec:disque_learning} to improve their robustness to CAFs. To demonstrate the usefulness of appearance mixing, we illustrate the outputs of EGTM when using the naive appearance replacement approach, also described in Section \ref{sec:disque_learning}. From Fig. \ref{fig:distm_confounding}, it may be observed that the green color of the wall in the source HDR, which is a CAF, results in a green sky in the predicted SDR when using appearance replacement. However, predicting the SDR image using appearance mixing significantly reduced the effect of the CAF, thereby improving prediction accuracy relative to the ``ground-truth'' target SDR. Note that in this example, both the example and target SDR images were generated using the Hable TMO \cite{ref:hable} with a desaturation parameter of 0. \highlight{A larger set of images tone-mapped using the proposed method is presented in the supplementary material.}

\highlight{To quantify the advantages of the novel appearance mixing method for EGTM, we analyzed the corresponding low-level image characteristics of ``ground-truth'' tone-mapped images and the corresponding predictions made by the trained EGTM model. For example, we sampled an HDR image from the dataset and tone-mapped it using the Hable TMO with a specific value of desaturation to obtain the ``ground-truth'' SDR image. Then, we created an example pair by applying the same procedure on a different HDR image sampled from the dataset. The example pair and the first HDR image were supplied to the trained EGTM model to predict the tone-mapped SDR image. Since the ``desaturation parameter'' of the Hable TMO controls color saturation (Sec \ref{sec:datasets_hdr}), we calculated the average value of the ``S'' channel of the HSV representations of both the ground-truth and predicted SDR images. We repeated this process at four distortion levels and for 200 different images. Finally, we calculated the linear correlation coefficient between the color saturations of the ground-truth and predicted images. To establish a baseline, we compared this against the existing ``appearance replacement'' strategy employed in the style-transfer literature (Sec. \ref{sec:disque_learning}). Furthermore, we carried out a similar procedure for various TMOs and tabulated the results in Table \ref{tab:egtm_val}. We also specify in Table \ref{tab:egtm_val} the low-level feature being varied (and therefore, being evaluated) by changing the parameters of each TMO. From these results, it may be inferred that the novel EGTM framework faithfully transfers the low-level appearance characteristics of the guide SDR images to the output, while the baseline appearance replacement method used in style transfer applications does not.}

\begin{table}[ht]
    \centering
    \caption{\highlight{Comparing the correlation between low-level image characteristics of ``ground-truth'' tone-mapped images and corresponding predictions made by Example-Guided Tone Mapping}}
    \label{tab:egtm_val}
    \begin{tabular}{|c|c|c|c|}
    \hline
        \multirow{2}{*}{\highlight{TMO}} & \multirow{2}{*}{\highlight{Characteristic}} & \highlight{App. Replacement} & \highlight{App. Mixing} \\
        & & \highlight{(Baseline)} & \highlight{(Ours)} \\
        \hline
        \highlight{Durand02} & \highlight{Contrast} & \highlight{0.4688} & \highlight{0.7931} \\
        \highlight{Eilertsen15} & \highlight{Contrast} & \highlight{0.4843} & \highlight{0.8289} \\
        \highlight{Hable} & \highlight{Color Saturation} & \highlight{0.4345} & \highlight{0.7451} \\
        \highlight{ITU21} & \highlight{Contrast} & \highlight{0.5615} & \highlight{0.9544} \\
        \highlight{Rana19} & \highlight{Color Saturation} & \highlight{0.5317} & \highlight{0.9153} \\
        \highlight{Reinhard02} & \highlight{Color Saturation} & \highlight{0.5031} & \highlight{0.8863} \\
        \highlight{Reinhard12} & \highlight{Color Hue} & \highlight{0.6561} & \highlight{0.7942} \\
        \highlight{Shan12} & \highlight{Contrast} & \highlight{0.3819} & \highlight{0.9136} \\
        \highlight{Yang21} & \highlight{Color Saturation} & \highlight{0.5266} & \highlight{0.9216} \\
        \hline
    \end{tabular}
\end{table}

\begin{figure}
    \centering
    \subfloat[\highlight{Blur}\label{fig:sdr_blur}]{%
      \includegraphics[width=0.9\linewidth]{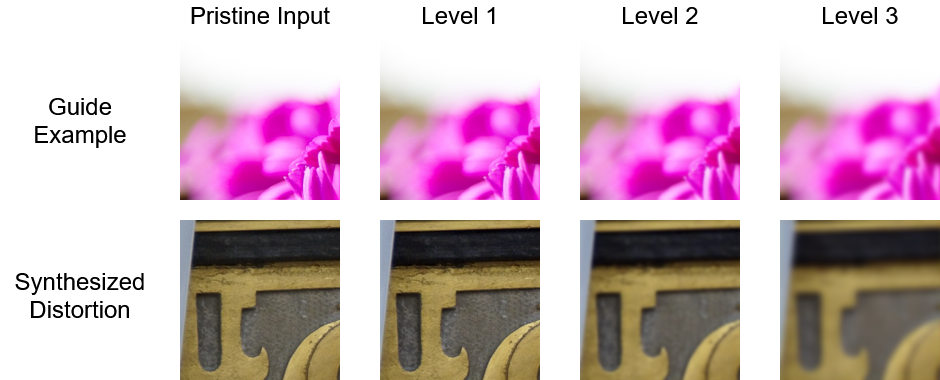}
    } \\
    \subfloat[\highlight{Darken}\label{fig:sdr_dark}]{%
      \includegraphics[width=0.9\linewidth]{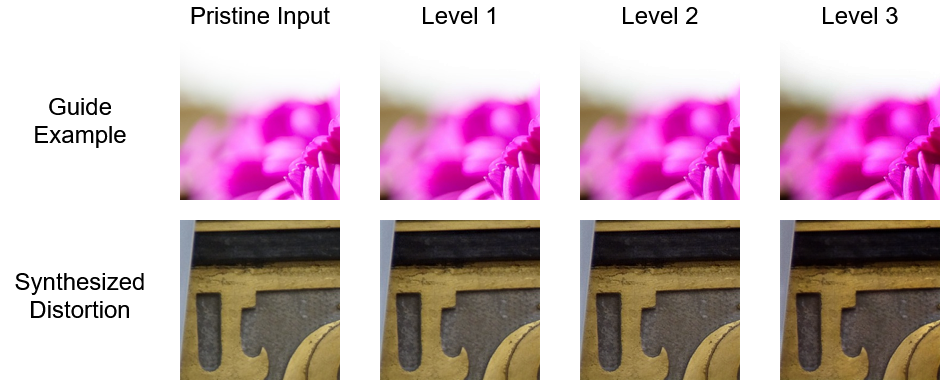}
    } \\
    \subfloat[\highlight{Noise}\label{fig:sdr_noise}]{%
      \includegraphics[width=0.9\linewidth]{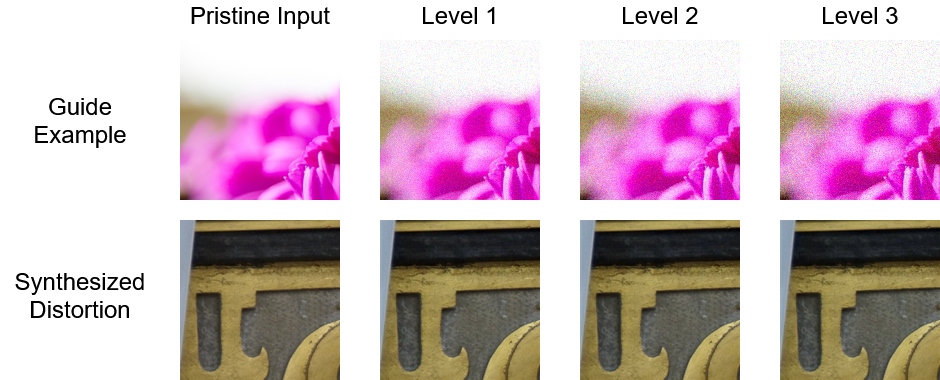}
    } \\
    \caption{\highlight{Synthesizing SDR distortions using guide examples. Blur and Darken distortions are successfully synthesized, while noise is not.}}
    \label{fig:dis_sdr}
\end{figure}

\highlight{Finally, we adopted a similar approach as EGTM to simulate distortions applied to SDR images, as described in Section \ref{sec:datasets_sdr}, which further demonstrates the ability of our trained models to perform example-guided image processing. In Figure \ref{fig:dis_sdr}, we show the results of simulating three distortions using paired examples: blur, darkening, and noise. From these figures, it may be seen that the model can synthesize spatially smooth blur and darkening distortions, though it is unable to generate stochastic distortions like noise. This observation, paired with the high quality prediction accuracy demonstrated in Table \ref{tab:sdr_accs}, shows that though the appearance encoder learns representations of all distortions, the decoder of the ResNet-50 architecture is unable to faithfully synthesize stochastic distortions. This limitation does not degrade tone-mapping quality since it does not typically include stochastic operations like adding noise. Further investigation into neural network architectures may bridge this gap.}

%% file: sections/discussion.tex
\section{\highlight{Discussion}}
\label{sec:discussion}
\highlight{Despite the promising performance of DisQUE and EGTM, the proposed DRL model has some limitations. First, since appearance mixing relies on differences between source and target domain appearance features, DisQUE is only able to perform full-reference quality modeling. Existing self-supervised methods such as CONTRIQUE and ReIQA can perform NR quality modeling since the appearance features, and not their differences, are used to distinguish between images. Combining these two methods may yield a network suitable for both FR and NR quality modeling.}

\highlight{Moreover, one of the assumptions used to design the DRL method was that appearance does not vary over image patches. However, appearance does vary over larger spatial regions, such as 1080p or 4K images. Therefore, the use of an average-pooled constant appearance vector, rather than a spatially varying feature map, limits the application of EGIP to patches. To enable the example-guided processing of high-resolution inputs, the appearance representation may be modified to include spatial information. Similarly, appearance characteristics may also temporally vary when processing video streams. Therefore, to enable example-guided video processing, the appearance representations may be modified to include temporal information, including necessary temporal consistency constraints.}

\highlight{Finally, we note that the EGTM results in Fig. \ref{fig:distm_confounding} showed that appearance mixing alone may not be sufficient for accurate EGTM. In this example, the predicted SDR was still a different shade of blue compared to the ground-truth target. This may be attributed to the fact that the example and input HDR images were significantly different in their visual characteristics - the example is predominantly a green wall, while the input is predominantly a blue sky. This suggests that a better approach during inference may be to use a bag of example pairs, and dynamically choose the most relevant example for every input image to be processed. More generally, a similar idea may be applied to the processing of high-resolution images by using patch attention methods to exploit ``good example patches'' from each example image pair.}

%% file: sections/conclusion.tex
\section{Conclusion}
\label{sec:conclusion}
We have developed a novel framework for disentangled representation learning using a new ``appearance mixing'' framework for adapting images across domains/appearance classes. The DRL network learns to decompose an input image into content and appearance-related features, which we used for two downstream tasks - perceptual quality modeling and example-guided image processing.

We found that our DRL-based quality model DisQUE achieved state-of-the-art accuracy when predicting the quality of tone-mapped and compressed HDR videos, and of synthetically distorted SDR images. In addition, we demonstrated the EGIP capabilities of the DRL model by performing example-guided tone mapping. Specifically, we gave examples of how \text{particular} image appearance features can be modulated using appropriately chosen examples.

%% file: main.bbl
\begin{thebibliography}{100}
\providecommand{\url}[1]{#1}
\csname url@samestyle\endcsname
\providecommand{\newblock}{\relax}
\providecommand{\bibinfo}[2]{#2}
\providecommand{\BIBentrySTDinterwordspacing}{\spaceskip=0pt\relax}
\providecommand{\BIBentryALTinterwordstretchfactor}{4}
\providecommand{\BIBentryALTinterwordspacing}{\spaceskip=\fontdimen2\font plus
\BIBentryALTinterwordstretchfactor\fontdimen3\font minus \fontdimen4\font\relax}
\providecommand{\BIBforeignlanguage}[2]{{%
\expandafter\ifx\csname l@#1\endcsname\relax
\typeout{** WARNING: IEEEtran.bst: No hyphenation pattern has been}%
\typeout{** loaded for the language `#1'. Using the pattern for}%
\typeout{** the default language instead.}%
\else
\language=\csname l@#1\endcsname
\fi
#2}}
\providecommand{\BIBdecl}{\relax}
\BIBdecl

\bibitem{ref:hdr_tv_survey}
\BIBentryALTinterwordspacing
CNET. Best {TV} for 2024: We tested {Samsung}, {LG}, {TCL}, {Vizio} and more. [Online]. Available: \url{https://www.cnet.com/tech/home-entertainment/best-tv/}
\BIBentrySTDinterwordspacing

\bibitem{ref:itu21}
ITU-R, ``{ITU-R BT.2446}: Methods for conversion of high dynamic range content to standard dynamic range content and vice-versa,'' 2021.

\bibitem{ref:hable}
\BIBentryALTinterwordspacing
J.~Hable. Uncharted 2: {HDR} lighting. [Online]. Available: \url{https://www.gdcvault.com/play/1012351/Uncharted-2-HDR}
\BIBentrySTDinterwordspacing

\bibitem{ref:dyn_optimizer}
\BIBentryALTinterwordspacing
I.~Katsavounidis, ``Dynamic optimizer - a perceptual video encoding optimization framework,'' March 2018. [Online]. Available: \url{https://netflixtechblog.com/dynamic-optimizer-a-perceptual-video-encoding-optimization-framework-e19f1e3a277f}
\BIBentrySTDinterwordspacing

\bibitem{ref:reinhard02}
E.~Reinhard, M.~Stark, P.~Shirley, and J.~Ferwerda, ``Photographic tone reproduction for digital images,'' \emph{ACM Transactions on Graphics}, vol.~21, no.~3, p. 267–276, Jul 2002.

\bibitem{ref:ssim}
Z.~Wang, A.~Bovik, H.~Sheikh, and E.~Simoncelli, ``Image quality assessment: from error visibility to structural similarity,'' \emph{IEEE Transactions on Image Processing}, vol.~13, no.~4, pp. 600--612, 2004.

\bibitem{ref:vif}
H.~R. Sheikh and A.~C. Bovik, ``Image information and visual quality,'' \emph{IEEE Transactions on Image Processing}, vol.~15, no.~2, pp. 430--444, 2006.

\bibitem{ref:strred}
R.~Soundararajan and A.~C. Bovik, ``Video quality assessment by reduced reference spatio-temporal entropic differencing,'' \emph{IEEE Transactions on Circuits and Systems for Video Technology}, vol.~23, no.~4, pp. 684--694, 2013.

\bibitem{ref:dlm}
S.~Li, F.~Zhang, L.~Ma, and K.~N. Ngan, ``Image quality assessment by separately evaluating detail losses and additive impairments,'' \emph{IEEE Transactions on Multimedia}, vol.~13, no.~5, pp. 935--949, 2011.

\bibitem{ref:vmaf}
Z.~Li, A.~Aaron, I.~Katsavounidis, A.~Moorthy, and M.~Manohara, ``Toward a practical perceptual video quality metric,'' \emph{The Netflix Tech Blog}, vol.~6, p.~2, 2016.

\bibitem{ref:funque}
A.~K. Venkataramanan, C.~Stejerean, and A.~C. Bovik, ``{FUNQUE}: Fusion of unified quality evaluators,'' in \emph{IEEE International Conference on Image Processing (ICIP)}, 2022, pp. 2147--2151.

\bibitem{ref:funque_plus}
A.~K. Venkataramanan, C.~Stejerean, I.~Katsavounidis, and A.~C. Bovik, ``One transform to compute them all: Efficient fusion-based full-reference video quality assessment,'' \emph{IEEE Transactions on Image Processing}, vol.~33, pp. 509--524, 2024.

\bibitem{ref:hdrmax_vmaf}
Z.~Shang, J.~P. Ebenezer, A.~K. Venkataramanan, Y.~Wu, H.~Wei, S.~Sethuraman, and A.~C. Bovik, ``A study of subjective and objective quality assessment of {HDR} videos,'' \emph{IEEE Transactions on Image Processing}, vol.~33, pp. 42--57, 2024.

\bibitem{ref:hdr_funque_plus}
A.~K. Venkataramanan, C.~Stejerean, I.~Katsavounidis, and A.~C. Bovik, ``A {FUNQUE} approach to the quality assessment of compressed {HDR} videos,'' \emph{arXiv preprint arXiv:2312.08524}, 2023.

\bibitem{ref:tmqi}
H.~Yeganeh and Z.~Wang, ``Objective quality assessment of tone-mapped images,'' \emph{IEEE Transactions on Image Processing}, vol.~22, no.~2, pp. 657--667, 2013.

\bibitem{ref:fsitm}
H.~{Ziaei Nafchi}, A.~{Shahkolaei}, R.~{Farrahi Moghaddam}, and M.~{Cheriet}, ``{FSITM}: A feature similarity index for tone-mapped images,'' \emph{IEEE Signal Processing Letters}, vol.~22, no.~8, pp. 1026--1029, 2015.

\bibitem{ref:cut_funque}
A.~K. Venkataramanan, C.~Stejerean, I.~Katsavounidis, H.~Tmar, and A.~C. Bovik, ``{Cut-FUNQUE}: Objective quality assessment of compressed and tone mapped high dynamic range videos,'' \emph{Manuscript Under Submission}, vol.~1, 2024.

\bibitem{ref:lpips}
R.~Zhang, P.~Isola, A.~A. Efros, E.~Shechtman, and O.~Wang, ``The unreasonable effectiveness of deep features as a perceptual metric,'' in \emph{IEEE/CVF International Conference on Computer Vision and Pattern Recognition (CVPR)}, June 2018.

\bibitem{ref:dists}
K.~Ding, K.~Ma, S.~Wang, and E.~P. Simoncelli, ``Image quality assessment: Unifying structure and texture similarity,'' \emph{IEEE Transactions on Pattern Analysis and Machine Intelligence}, vol.~44, no.~5, pp. 2567--2581, 2022.

\bibitem{ref:deep_wsd}
X.~Liao, B.~Chen, H.~Zhu, S.~Wang, M.~Zhou, and S.~Kwong, ``{DeepWSD}: Projecting degradations in perceptual space to wasserstein distance in deep feature space,'' in \emph{ACM International Conference on Multimedia}, 2022, p. 970–978.

\bibitem{ref:brisque}
A.~Mittal, A.~K. Moorthy, and A.~C. Bovik, ``No-reference image quality assessment in the spatial domain,'' \emph{IEEE Transactions on Image Processing}, vol.~21, no.~12, pp. 4695--4708, 2012.

\bibitem{ref:diivine}
A.~K. Moorthy and A.~C. Bovik, ``Blind image quality assessment: From natural scene statistics to perceptual quality,'' \emph{IEEE Transactions on Image Processing}, vol.~20, no.~12, pp. 3350--3364, 2011.

\bibitem{ref:tlvqm}
J.~Korhonen, ``Two-level approach for no-reference consumer video quality assessment,'' \emph{IEEE Transactions on Image Processing}, vol.~28, no.~12, pp. 5923--5938, 2019.

\bibitem{ref:higrade}
D.~Kundu, D.~Ghadiyaram, A.~C. Bovik, and B.~Evans, ``No-reference quality assessment of tone-mapped {HDR} pictures,'' \emph{IEEE Transactions on Image Processing}, vol.~26, no.~6, pp. 2957--2971, 2017.

\bibitem{ref:chipqa}
J.~P. Ebenezer, Z.~Shang, Y.~Wu, H.~Wei, S.~Sethuraman, and A.~C. Bovik, ``{ChipQA}: No-reference video quality prediction via space-time chips,'' \emph{IEEE Transactions on Image Processing}, vol.~30, pp. 8059--8074, 2021.

\bibitem{ref:paq2piq}
Z.~Ying, H.~Niu, P.~Gupta, D.~Mahajan, D.~Ghadiyaram, and A.~Bovik, ``From patches to pictures {(PaQ-2-PiQ)}: Mapping the perceptual space of picture quality,'' in \emph{IEEE/CVF International Conference on Computer Vision and Pattern Recognition}, 2020, pp. 3575--3585.

\bibitem{ref:patchvq}
Z.~Ying, M.~Mandal, D.~Ghadiyaram, and A.~Bovik, ``{Patch-VQ}: `{Patching} up' the video quality problem,'' in \emph{IEEE/CVF International Conference on Computer Vision and Pattern Recognition}, 2021, pp. 14\,019--14\,029.

\bibitem{ref:qfm_iqm}
X.~Li, T.~Gao, X.~Zheng, R.~Hu, J.~Zheng, Y.~Shen, K.~Li, Y.~Liu, P.~Dai, Y.~Zhang \emph{et~al.}, ``Adaptive feature selection for no-reference image quality assessment using contrastive mitigating semantic noise sensitivity,'' \emph{arXiv preprint arXiv:2312.06158}, 2023.

\bibitem{ref:musiq}
J.~Ke, Q.~Wang, Y.~Wang, P.~Milanfar, and F.~Yang, ``{MUSIQ}: Multi-scale image quality transformer,'' in \emph{IEEE/CVF International Conference on Computer Vision (ICCV)}, October 2021, pp. 5148--5157.

\bibitem{ref:rkiqt}
X.~Li, J.~Zheng, X.~Zheng, R.~Hu, E.~Zhang, Y.~Gao, Y.~Shen, K.~Li, Y.~Liu, P.~Dai \emph{et~al.}, ``Less is more: Learning reference knowledge using no-reference image quality assessment,'' \emph{arXiv preprint arXiv:2312.00591}, 2023.

\bibitem{ref:loda}
K.~Xu, L.~Liao, J.~Xiao, C.~Chen, H.~Wu, Q.~Yan, and W.~Lin, ``Local distortion aware efficient transformer adaptation for image quality assessment,'' \emph{arXiv preprint arXiv:2308.12001}, 2023.

\bibitem{ref:contrique}
P.~C. Madhusudana, N.~Birkbeck, Y.~Wang, B.~Adsumilli, and A.~C. Bovik, ``Image quality assessment using contrastive learning,'' \emph{IEEE Transactions on Image Processing}, vol.~31, pp. 4149--4161, 2022.

\bibitem{ref:reiqa}
A.~Saha, S.~Mishra, and A.~C. Bovik, ``{Re-IQA}: Unsupervised learning for image quality assessment in the wild,'' in \emph{IEEE/CVF International Conference on Computer Vision and Pattern Recognition}, 2023, pp. 5846--5855.

\bibitem{ref:conviqt}
P.~C. Madhusudana, N.~Birkbeck, Y.~Wang, B.~Adsumilli, and A.~C. Bovik, ``{CONVIQT}: Contrastive video quality estimator,'' \emph{IEEE Transactions on Image Processing}, vol.~32, pp. 5138--5152, 2023.

\bibitem{ref:simclr}
T.~Chen, S.~Kornblith, M.~Norouzi, and G.~Hinton, ``A simple framework for contrastive learning of visual representations,'' in \emph{International Conference on Machine Learning}, 2020, pp. 1597--1607.

\bibitem{ref:moco}
K.~He, H.~Fan, Y.~Wu, S.~Xie, and R.~Girshick, ``Momentum contrast for unsupervised visual representation learning,'' in \emph{IEEE/CVF International Conference on Computer Vision and Pattern Recognition}, 2020, pp. 9729--9738.

\bibitem{ref:mocov2}
X.~Chen, H.~Fan, R.~Girshick, and K.~He, ``Improved baselines with momentum contrastive learning,'' \emph{arXiv preprint arXiv:2003.04297}, 2020.

\bibitem{ref:drl_survey}
X.~Wang, H.~Chen, S.~Tang, Z.~Wu, and W.~Zhu, ``Disentangled representation learning,'' \emph{arXiv preprint arXiv:2211.11695}, 2022.

\bibitem{ref:factor_vae}
H.~Kim and A.~Mnih, ``Disentangling by factorising,'' in \emph{International Conference on Machine Learning}, Jul 2018, pp. 2649--2658.

\bibitem{ref:beta_tcvae}
R.~T. Chen, X.~Li, R.~B. Grosse, and D.~K. Duvenaud, ``Isolating sources of disentanglement in variational autoencoders,'' \emph{Advances in Neural Information Processing Systems}, vol.~31, 2018.

\bibitem{ref:info_gan}
X.~Chen, Y.~Duan, R.~Houthooft, J.~Schulman, I.~Sutskever, and P.~Abbeel, ``{InfoGAN}: Interpretable representation learning by information maximizing generative adversarial nets,'' in \emph{Advances in Neural Information Processing Systems}, vol.~29, 2016.

\bibitem{ref:ps_sc_gan}
X.~Zhu, C.~Xu, and D.~Tao, ``Where and what? examining interpretable disentangled representations,'' in \emph{IEEE/CVF International Conference on Computer Vision and Pattern Recognition (CVPR)}, June 2021, pp. 5861--5870.

\bibitem{ref:orojar}
Y.~Wei, Y.~Shi, X.~Liu, Z.~Ji, Y.~Gao, Z.~Wu, and W.~Zuo, ``Orthogonal jacobian regularization for unsupervised disentanglement in image generation,'' in \emph{IEEE/CVF International Conference on Computer Vision (ICCV)}, October 2021, pp. 6721--6730.

\bibitem{ref:barlow}
J.~Zbontar, L.~Jing, I.~Misra, Y.~LeCun, and S.~Deny, ``Barlow twins: Self-supervised learning via redundancy reduction,'' in \emph{International Conference on Machine Learning}, 2021, pp. 12\,310--12\,320.

\bibitem{ref:dr_gan}
L.~Tran, X.~Yin, and X.~Liu, ``Disentangled representation learning {GAN} for pose-invariant face recognition,'' in \emph{IEEE/CVF International Conference on Computer Vision and Pattern Recognition (CVPR)}, Jul 2017.

\bibitem{ref:map_ivr}
L.~Liu, J.~Li, L.~Niu, R.~Xu, and L.~Zhang, ``Activity image-to-video retrieval by disentangling appearance and motion,'' in \emph{{AAAI} Conference on Artificial Intelligence}, vol.~35, no.~3, 2021, pp. 2145--2153.

\bibitem{ref:club_mi}
P.~Cheng, W.~Hao, S.~Dai, J.~Liu, Z.~Gan, and L.~Carin, ``{CLUB}: A contrastive log-ratio upper bound of mutual information,'' in \emph{International Conference on Machine Learning}, vol. 119, 13--18 Jul 2020, pp. 1779--1788.

\bibitem{ref:adv_disrep}
E.~H. Sanchez, M.~Serrurier, and M.~Ortner, ``Learning disentangled representations via mutual information estimation,'' in \emph{European Conference on Computer Vision (ECCV)}, 2020, pp. 205--221.

\bibitem{ref:drit}
H.-Y. Lee, H.-Y. Tseng, J.-B. Huang, M.~Singh, and M.-H. Yang, ``Diverse image-to-image translation via disentangled representations,'' in \emph{European Conference on Computer Vision (ECCV)}, September 2018.

\bibitem{ref:disrep_domain}
V.-H. Tran and C.-C. Huang, ``Domain adaptation meets disentangled representation learning and style transfer,'' in \emph{IEEE International Conference on Systems, Man and Cybernetics (SMC)}, 2019, pp. 2998--3005.

\bibitem{ref:qairn}
H.~E. Gedik, A.~K. Venkataramanan, and A.~C. Bovik, ``Joint deep image restoration and unsupervised quality assessment,'' \emph{arXiv preprint arXiv:2311.16372}, 2023.

\bibitem{ref:qdnet}
L.~Wang, Q.~Wu, K.~N. Ngan, H.~Li, F.~Meng, and L.~Xu, ``Blind tone-mapped image quality assessment and enhancement via disentangled representation learning,'' in \emph{Asia-Pacific Signal and Information Processing Association Annual Summit and Conference (APSIPA ASC)}, 2020, pp. 1096--1102.

\bibitem{ref:driqa}
Z.~Ye, Y.~Wu, D.~Liao, T.~Yu, J.~Yang, and J.~Hu, ``{DRIQA-NR}: No-reference image quality assessment based on disentangled representation,'' \emph{Signal, Image and Video Processing}, vol.~17, no.~3, pp. 661--669, 2023.

\bibitem{ref:siamese}
J.~Bromley, I.~Guyon, Y.~LeCun, E.~S\"{a}ckinger, and R.~Shah, ``Signature verification using a {``Siamese''} time delay neural network,'' in \emph{Advances in Neural Information Processing Systems}, vol.~6, 1993.

\bibitem{ref:intrinsics}
H.~Barrow, J.~Tenenbaum, A.~Hanson, and E.~Riseman, ``Recovering intrinsic scene characteristics,'' \emph{Computer Vision Systems}, vol.~2, no. 3-26, p.~2, 1978.

\bibitem{ref:maxim}
Z.~Tu, H.~Talebi, H.~Zhang, F.~Yang, P.~Milanfar, A.~C. Bovik, and Y.~Li, ``{MAXIM}: Multi-axis {MLP} for image processing,'' in \emph{IEEE/CVF International Conference on Computer Vision and Pattern Recognition (CVPR)}, June 2022, pp. 5769--5780.

\bibitem{ref:infonce}
A.~v.~d. Oord, Y.~Li, and O.~Vinyals, ``Representation learning with contrastive predictive coding,'' \emph{arXiv preprint arXiv:1807.03748}, 2018.

\bibitem{ref:mocov3}
X.~Chen, S.~Xie, and K.~He, ``An empirical study of training self-supervised vision transformers,'' in \emph{IEEE/CVF International Conference on Computer Vision}, 2021, pp. 9640--9649.

\bibitem{ref:unet}
O.~Ronneberger, P.~Fischer, and T.~Brox, ``U-net: Convolutional networks for biomedical image segmentation,'' in \emph{International Conference on Medical Image Computing and Computer-Assisted Intervention, Munich, Germany, October 5-9}, 2015, pp. 234--241.

\bibitem{ref:enhanced_resnet}
B.~Lim, S.~Son, H.~Kim, S.~Nah, and K.~Mu~Lee, ``Enhanced deep residual networks for single image super-resolution,'' in \emph{IEEE/CVF International Conference on Computer Vision and Pattern Recognition (CVPR) Workshops}, Jul 2017.

\bibitem{ref:dynamic_deblur}
S.~Nah, T.~Hyun~Kim, and K.~Mu~Lee, ``Deep multi-scale convolutional neural network for dynamic scene deblurring,'' in \emph{IEEE/CVF International Conference on Computer Vision and Pattern Recognition (CVPR)}, Jul 2017, pp. 3883--3891.

\bibitem{ref:instance_norm}
D.~Ulyanov, A.~Vedaldi, and V.~Lempitsky, ``Improved texture networks: Maximizing quality and diversity in feed-forward stylization and texture synthesis,'' in \emph{IEEE/CVF International Conference on Computer Vision and Pattern Recognition (CVPR)}, 2017, pp. 4105--4113.

\bibitem{ref:condin}
V.~Dumoulin, J.~Shlens, and M.~Kudlur, ``A learned representation for artistic style,'' \emph{arXiv preprint arXiv:1610.07629}, 2016.

\bibitem{ref:ranet}
F.~Wang, M.~Jiang, C.~Qian, S.~Yang, C.~Li, H.~Zhang, X.~Wang, and X.~Tang, ``Residual attention network for image classification,'' in \emph{IEEE/CVF International Conference on Computer Vision and Pattern Recognition (CVPR)}, Jul 2017, pp. 3156--3164.

\bibitem{ref:adain}
X.~Huang and S.~Belongie, ``Arbitrary style transfer in real-time with adaptive instance normalization,'' in \emph{IEEE/CVF International Conference on Computer Vision (ICCV)}, Oct 2017, pp. 1501--1510.

\bibitem{ref:msml}
Q.~He, D.~Li, T.~Jiang, and M.~Jiang, ``Quality assessment for tone-mapped {HDR} images using multi-scale and multi-layer information,'' in \emph{IEEE International Conference on Multimedia and Expo Workshops}, 2018, pp. 1--6.

\bibitem{ref:essim}
A.~K. Venkataramanan, C.~Wu, A.~C. Bovik, I.~Katsavounidis, and Z.~Shahid, ``A hitchhiker’s guide to structural similarity,'' \emph{IEEE Access}, vol.~9, pp. 28\,872--28\,896, 2021.

\bibitem{ref:gmsd}
W.~Xue, L.~Zhang, X.~Mou, and A.~C. Bovik, ``Gradient magnitude similarity deviation: A highly efficient perceptual image quality index,'' \emph{IEEE Transactions on Image Processing}, vol.~23, no.~2, pp. 684--695, 2013.

\bibitem{ref:live_ugc_hdr}
\BIBentryALTinterwordspacing
S.~Saini, A.~K. Venkataramanan, and A.~C. Bovik, ``The {LIVE} user-generated {HDR} video dataset,'' 2024. [Online]. Available: \url{https://live.ece.utexas.edu/research/LIVE_UGC_HDR/index.html}
\BIBentrySTDinterwordspacing

\bibitem{ref:hlg}
T.~Borer and A.~Cotton, ``A display-independent high dynamic range television system,'' \emph{SMPTE Motion Imaging Journal}, vol. 125, no.~4, pp. 50--56, 2016.

\bibitem{ref:pq}
{SMPTE}, ``High dynamic range electro-optical transfer function of mastering reference displays,'' \emph{SMPTE Standard}, vol. 2084, p.~11, 2014.

\bibitem{ref:durand02}
F.~Durand and J.~Dorsey, ``Fast bilateral filtering for the display of high-dynamic-range images,'' in \emph{ACM Annual Conference on Computer Graphics and Interactive Techniques}, 2002, p. 257–266.

\bibitem{ref:shan12}
Q.~Shan, T.~DeRose, and J.~Anderson, ``Tone mapping high dynamic range videos using wavelets,'' \emph{Pixar Technical Memo}, 2012.

\bibitem{ref:swt}
G.~P. Nason and B.~W. Silverman, \emph{The Stationary Wavelet Transform and some Statistical Applications}.\hskip 1em plus 0.5em minus 0.4em\relax Springer New York, 1995, pp. 281--299.

\bibitem{ref:reinhard12}
E.~Reinhard, T.~Pouli, T.~Kunkel, B.~Long, A.~Ballestad, and G.~Damberg, ``Calibrated image appearance reproduction,'' \emph{ACM Trans. Graph.}, vol.~31, no.~6, Nov 2012.

\bibitem{ref:eilertsen15}
G.~Eilertsen, R.~K. Mantiuk, and J.~Unger, ``Real-time noise-aware tone mapping,'' \emph{ACM Trans. Graph.}, vol.~34, no.~6, Nov 2015.

\bibitem{ref:oskarsson17}
M.~Oskarsson, ``Temporally consistent tone mapping of images and video using optimal k-means clustering,'' \emph{Journal of Mathematical Imaging and Vision}, vol.~57, no.~2, pp. 225--238, Feb 2017.

\bibitem{ref:rana19}
A.~Rana, P.~Singh, G.~Valenzise, F.~Dufaux, N.~Komodakis, and A.~Smolic, ``Deep tone mapping operator for high dynamic range images,'' \emph{IEEE Transactions on Image Processing}, vol.~29, pp. 1285--1298, 2020.

\bibitem{ref:yang21}
J.~Yang, Z.~Liu, M.~Lin, S.~Yanushkevich, and O.~Yadid-Pecht, ``Deep reformulated laplacian tone mapping,'' \emph{arXiv preprint arXiv:2102.00348}, 2021.

\bibitem{ref:live_tmhdr}
A.~K. Venkataramanan and A.~C. Bovik, ``Subjective quality assessment of compressed tone-mapped high dynamic range videos,'' \emph{Manuscript Under Preparation}, vol.~1, 2024.

\bibitem{ref:x264}
\BIBentryALTinterwordspacing
{VideoLAN}, ``x264.'' [Online]. Available: \url{https://code.videolan.org/videolan/x264.git}
\BIBentrySTDinterwordspacing

\bibitem{ref:dolbyvision}
\BIBentryALTinterwordspacing
(2016) An introduction to {Dolby Vision}. [Online]. Available: \url{https://professional.dolby.com/siteassets/pdfs/dolby-vision-whitepaper_an-introduction-to-dolby-vision_0916.pdf}
\BIBentrySTDinterwordspacing

\bibitem{ref:kadis700k}
H.~Lin, V.~Hosu, and D.~Saupe, ``{DeepFL-IQA}: Weak supervision for deep {IQA} feature learning,'' \emph{arXiv preprint arXiv:2001.08113}, 2020.

\bibitem{ref:ava}
N.~Murray, L.~Marchesotti, and F.~Perronnin, ``{AVA}: A large-scale database for aesthetic visual analysis,'' in \emph{IEEE/CVF International Conference on Computer Vision and Pattern Recognition}, 2012, pp. 2408--2415.

\bibitem{ref:blur}
E.~Mavridaki and V.~Mezaris, ``No-reference blur assessment in natural images using {Fourier} transform and spatial pyramids,'' in \emph{IEEE International Conference on Image Processing (ICIP)}, 2014, pp. 566--570.

\bibitem{ref:voc}
M.~Everingham, L.~Van~Gool, C.~K. Williams, J.~Winn, and A.~Zisserman, ``The {Pascal} visual object classes {(VOC)} challenge,'' \emph{International Journal of Computer Vision}, vol.~88, pp. 303--338, 2010.

\bibitem{ref:coco}
T.-Y. Lin, M.~Maire, S.~Belongie, J.~Hays, P.~Perona, D.~Ramanan, P.~Doll{\'a}r, and C.~L. Zitnick, ``Microsoft {COCO}: Common objects in context,'' in \emph{European Conference on Computer Vision}, 2014, pp. 740--755.

\bibitem{ref:places}
B.~Zhou, A.~Lapedriza, J.~Xiao, A.~Torralba, and A.~Oliva, ``Learning deep features for scene recognition using places database,'' in \emph{Advances in Neural Information Processing Systems}, vol.~27, 2014.

\bibitem{ref:live_iqa}
H.~R. Sheikh, M.~F. Sabir, and A.~C. Bovik, ``A statistical evaluation of recent full reference image quality assessment algorithms,'' \emph{IEEE Transactions on Image Processing}, vol.~15, no.~11, pp. 3440--3451, 2006.

\bibitem{ref:csiq}
E.~C. Larson and D.~M. Chandler, ``Most apparent distortion: full-reference image quality assessment and the role of strategy,'' \emph{Journal of Electronic Imaging}, vol.~19, no.~1, pp. 011\,006--011\,006, 2010.

\bibitem{ref:tid2013}
N.~Ponomarenko, O.~Ieremeiev, V.~Lukin, K.~Egiazarian, L.~Jin, J.~Astola, B.~Vozel, K.~Chehdi, M.~Carli, F.~Battisti \emph{et~al.}, ``Color image database {TID2013}: Peculiarities and preliminary results,'' in \emph{European Workshop on Visual Information Processing (EUVIP)}, 2013, pp. 106--111.

\bibitem{ref:kadid10k}
H.~Lin, V.~Hosu, and D.~Saupe, ``{KADID-10k}: A large-scale artificially distorted {IQA} database,'' in \emph{International Conference on Quality of Multimedia Experience (QoMEX)}, 2019, pp. 1--3.

\bibitem{ref:fsim}
L.~Zhang, L.~Zhang, X.~Mou, and D.~Zhang, ``{FSIM}: A feature similarity index for image quality assessment,'' \emph{IEEE Transactions on Image Processing}, vol.~20, no.~8, pp. 2378--2386, 2011.

\bibitem{ref:cornia}
P.~Ye, J.~Kumar, L.~Kang, and D.~Doermann, ``Unsupervised feature learning framework for no-reference image quality assessment,'' in \emph{IEEE/CVF International Conference on Computer Vision and Pattern Recognition}, 2012, pp. 1098--1105.

\bibitem{ref:dbcnn}
W.~Zhang, K.~Ma, J.~Yan, D.~Deng, and Z.~Wang, ``Blind image quality assessment using a deep bilinear convolutional neural network,'' \emph{IEEE Transactions on Circuits and Systems for Video Technology}, vol.~30, no.~1, pp. 36--47, 2020.

\bibitem{ref:pqr}
H.~Zeng, L.~Zhang, and A.~C. Bovik, ``A probabilistic quality representation approach to deep blind image quality prediction,'' \emph{arXiv preprint arXiv:1708.08190}, 2017.

\bibitem{ref:hyperiqa}
S.~Su, Q.~Yan, Y.~Zhu, C.~Zhang, X.~Ge, J.~Sun, and Y.~Zhang, ``Blindly assess image quality in the wild guided by a self-adaptive hyper network,'' in \emph{IEEE/CVF International Conference on Computer Vision and Pattern Recognition (CVPR)}, June 2020, pp. 3667--3676.

\bibitem{ref:drf_iqa}
W.~Kim, A.-D. Nguyen, S.~Lee, and A.~C. Bovik, ``Dynamic receptive field generation for full-reference image quality assessment,'' \emph{IEEE Transactions on Image Processing}, vol.~29, pp. 4219--4231, 2020.

\bibitem{ref:satqa}
J.~Shi, P.~Gao, and J.~Qin, ``Transformer-based no-reference image quality assessment via supervised contrastive learning,'' \emph{arXiv preprint arXiv:2312.06995}, 2023.

\bibitem{ref:guided_filtering}
K.~He, J.~Sun, and X.~Tang, ``Guided image filtering,'' \emph{IEEE Transactions on Pattern Analysis and Machine Intelligence}, vol.~35, no.~6, pp. 1397--1409, 2013.

\end{thebibliography}
